\documentclass[
    reprint,
    amsfonts,
    amssymb,
    amsmath,
    showkeys,
    pra,
    nofootinbib,
    superscriptaddress,
    twocolumn,
    longbibliography,
    aps
]{revtex4-2}

\usepackage{array}[=2019/01/01]

\usepackage[english]{babel}
\usepackage[utf8]{inputenc}
\usepackage[T1]{fontenc}
\usepackage[most]{tcolorbox} 
\usepackage{enumitem}        
\usepackage{amsmath,amssymb,amsfonts,mathtools}
\usepackage{bm, svg}      
\usepackage{bbm}     
\usepackage{braket}  
\usepackage{physics} 
\usepackage{amsthm, dsfont}  
\usepackage{soul} 

\usepackage{graphicx,tabularx}
\usepackage[caption=false,font=small]{subfig}
\usepackage{booktabs}
\usepackage{placeins}   

\usepackage[most]{tcolorbox} 
\newtcolorbox[auto counter]{myalgobox}[2]{
  colback=white,
  colframe=gray!60!black,
  colbacktitle=gray!25,
  coltitle=black,
  title={Box~\thetcbcounter: #1},
  label={#2},
  boxrule=0.8pt,
  sharp corners,
  enhanced,
  attach boxed title to top center={yshift=-2mm},
  boxed title style={boxrule=0.5pt, colframe=gray!60!black},
  breakable 
}
\usepackage{xcolor}
\usepackage{enumitem}
\usepackage{csquotes}
\usepackage{tikz}
\usepackage[colorlinks=true,citecolor=blue,linkcolor=magenta]{hyperref}
\DeclareMathOperator{\Vect}{vec}

\begin{document}
\preprint{APS/123-QED}

\title{Strategy optimization for Bayesian quantum parameter estimation with finite copies: Adaptive greedy, parallel, sequential, and general strategies}
\author{Erik L. André}
\email{erik.andre@tuwien.ac.at}
\affiliation{Technische Universität Wien, Atominstitut, Vienna Center for Quantum Science and Technology, Stadionallee 2, 1020 Vienna, Austria}
\author{Jessica Bavaresco}
\email{jessica.bavaresco@lip6.fr}
\affiliation{Sorbonne Université, CNRS, LIP6, F-75005 Paris, France}
\author{Mohammad Mehboudi}
\thanks{Corresponding Author}
\email{mohammad.mehboudi@gmail.com}
\affiliation{Technische Universität Wien, Atominstitut, Vienna Center for Quantum Science and Technology, Stadionallee 2, 1020 Vienna, Austria}

\date{\today}

\begin{abstract}
In this work, we study Bayesian quantum parameter estimation given a finite number of uses of the process encoding one or more unknown physical quantities. We analyze the performance of different kinds of quantum metrological protocols, including the conventional parallel, sequential, or general, which may involve an indefinite causal order. We also analyze a class of hybrid adaptive \textit{greedy} strategies, which are based on classical feedforward between quantum protocols to optimize the next round.
Within each class, the central question is to determine the optimal strategy---namely, the choice of optimal input state, control operations, measurement, and estimators. Using the formalism of higher-order operations, we develop an algorithm that searches for the optimal solution, and we provide a numerical implementation based on semidefinite programming. Our benchmark cases, specifically those against existing analytical solutions, demonstrate how powerful and precise our method is. We further demonstrate the strength of our algorithm in several examples, from single to multiparameter estimation, and with various prior distributions. Particularly, we find examples where the adaptive greedy strategy matches the performance of general strategies with indefinite causal order, and at the same time examples that showcase a strict hierarchy between all different classes.
\end{abstract}
\maketitle
\section{Introduction}
The Bayesian approach to metrology is a very successful method in estimating unknown parameters of physical systems~\cite{RevModPhys.83.943,Trotta01032008,Thrane_Talbot_2019}. In quantum systems specifically, it has the potential to advise on how to design the probing protocol as well, namely, how to prepare the optimal quantum state for the probe and the optimal measurement on the probe after acquiring information about the unknown parameter(s)~\cite{helstrom1969quantum,1054643,e20090628,Morelli_2021,PhysRevA.101.032114}. Particularly when dealing with finite data~\cite{PhysRevResearch.6.023305,qbn1-p6bq,rubio_thesis}, it can advise us how to generate and optimally use such data. Nonetheless, except in specific cases~\cite{1054643,PhysRevA.83.061802,Macieszczak_2014,Rubio_2023} where the Bayesian cost function is simple enough---such as quadratic distance---analytical solutions to the problem do not exist. 
Recently, there has been an extensive focus in coming up with algorithms that efficiently find the optimal protocol numerically~\cite{PhysRevResearch.6.023305,qbn1-p6bq,PRXQuantum.4.020333}. Specifically, the approach proposed in \cite{PhysRevResearch.6.023305} allows one to find the optimal single-shot strategy approximately, with arbitrary precision, being also suitable for multiparameter estimation.

On a different front, talking about parameter estimation with quantum systems always brings up the subject of quantum advantages~\cite{PhysRevLett.72.3439,PhysRevLett.96.010401,Giovannetti2011,doi:10.1142/S0219749909004839,Toth_2014}. That is, when having access to a fixed number of calls to the process that encodes the parameter (a quantum channel), how can quantum resources assist to better estimate the set of parameters. One often categorizes causally ordered quantum-assisted protocols into two main classes: parallel (PAR) and sequential (SEQ) strategies~\cite{PhysRevLett.96.010401,PhysRevLett.98.090501}, which we will define rigorously later, with quantum or classical memory~\cite{Ohst2026characterising}. Beyond this, indefinite causal order (ICO) protocols~\cite{hardy2005probability,PhysRevA.88.022318} are also explored for potential quantum advantages; however, quantum realizations of these processes are generally unknown.
The comparison between the different protocols, and whether there is a strict hierarchy between them, is a topic that is well studied beyond quantum metrology, for instance, in quantum channel discrimination~\cite{PhysRevA.86.040301,PhysRevLett.127.200504,Bavaresco2022}, quantum communication~\cite{PhysRevLett.120.120502,PhysRevA.92.052326} and computation~\cite{PhysRevLett.128.230503,PhysRevLett.113.250402}. The study of parallel, sequential, and ICO strategies for quantum metrology has mainly been carried out in the asymptotic limit of large data with the frequentist approach, namely, when the quantum Fisher information (QFI) is a suitable figure of merit~\cite{PhysRevLett.124.190503,PhysRevLett.130.070803,Yin2023,Kurdzialek_2025}. 

In a Bayesian formalism with a general cost function, performing such a comparison has not been possible yet, due to the difficulty in finding the optimal solution in each class. This is the gap that we intend to bridge here. Particularly, we develop the necessary tools and provide their numerical implementation, to find the optimal strategies for parallel, sequential, and indefinite causal order protocols. We also consider adaptive \textit{greedy} protocols as a viable solution for scenarios with limited or no access to quantum memory. A priori, one knows that the class of sequential strategies contains, as particular cases, both parallel strategies and adaptive greedy strategies with classical memory, while adaptive greedy and parallel protocols are not subsets of one another. At the same time, the class of general strategies, which may exhibit an indefinite causal order, contains sequential strategies as a particular case.

We use our methods to demonstrate via several examples that these hierarchies can be strict in certain Bayesian metrology protocols. Furthermore, we also find examples in which all protocols perform equally well.

The remainder of the paper is structured as follows. In Section~\ref{sec:framework} we define a generic Bayesian estimation problem which also sets up our notation. In Section~\ref{sec:protocols} we discuss different classes of protocols for parameter estimation, namely adaptive greedy, parallel, sequential, and indefinite causal order. In Section~\ref{sec:numerical_impl} we present the algorithms that enable an efficient numerical approximate solution to the different protocols. In Section~\ref{sec:examples} we exploit our formalism to construct the optimal protocols in several examples, from single to multiparameter estimation, from unitary to dissipative encoding, and with various prior distributions. In simple cases with specific symmetries where analytical answers exist, our numerical results predict exactly the same solutions, demonstrating the optimality of our strategies, as well as the strength of our methods and quality of our approximations.
Finally, in Section~\ref{sec:discussion} we conclude by discussing potential future directions. 

\section{Framework: Single-shot scenario}\label{sec:framework}
We start with the basic setting, i.e., a single-shot scenario. Multiple-shot scenarios, which are the main subject of this work, will be discussed in the next section. Bayes' rule is used to update one's belief about a set of $q$ unknown parameters $\bm{\theta} \in \mathbb{R}^q$, captured by the \textit{prior} probability density $p(\bm{\theta})$ conditioned on new evidence (observation) $i$ to obtain the \textit{posterior} distribution $p({\bm \theta}|i)$~\cite{Berger1985,Bernardo1994,RevModPhys.83.943}
\begin{align}\label{bayesrule}
    p({\bm \theta} | i) = \frac{p(i|{\bm \theta})p({\bm \theta})}{p(i)},
\end{align}
where $p(i|{\bm \theta}) $ is the probability of observing outcome $i$ conditioned on the parameters being $\bm{\theta}$, and with $p(i) = \int \dd{\bm \theta} p(i|{\bm \theta})p({\bm \theta})$ being the marginal likelihood. 
The Bayesian formalism is a strong approach to metrology, as it brings together both the prior knowledge and the new evidence together to assign an estimate ${{\bm{\hat \theta}}}_i \in \mathbb{R}^q$ to the unknown parameters ${\bm \theta}$. 
The quality of the estimation is often quantified by using a cost function $c({\bm \theta},{\bm {\hat \theta}}_i)$ that needs to be chosen thoughtfully depending on the problem. The choice of the cost function is also determining in designing the optimal estimation procedure. 

In quantum physics, the parameters are encoded via completely positive and trace-preserving (CPTP) channels $\Lambda_{{\bm \theta}}: {\cal L}({\cal H}_{\rm I}) \mapsto {\cal L}({\cal H}_{\rm O})$---which we also refer to as quantum channels~\cite{Kraus1971,Nielsen2012,Wilde2016}. Here, ${\cal L}({\cal H}_{\rm I})$ and ${\cal L}({\cal H}_{\rm O})$ represent the set of operators acting on the input Hilbert space ${\cal H}_{\rm I}$ with dimension $d_{\rm I}$ and the output Hilbert space ${\cal H}_{\rm O}$ with dimension $d_{\rm O}$, respectively. Often, it is advantageous to consider probes (inputs to the channel) that are correlated to auxiliary systems that do not undergo the action of the CPTP map. Denoting their joint density operator with $\rho \in {\cal L}({\cal H}_{\rm I} \otimes {\cal H}_{\rm aux})$, the output reads $\rho_{{\bm \theta}} \coloneqq (\Lambda_{{\bm \theta}}\otimes \mathds{1}) (\rho) \in {\cal L}({\cal H}_{\rm O} \otimes {\cal H}_{\rm aux})$, where the auxiliary Hilbert space ${\cal H}_{\rm aux}$ has dimension $d_{\rm I}$. Furthermore, the label $i$ refers to the different outcomes of a measurement represented by a positive operator-valued measure (POVM) $M \coloneqq \{M_i\}_i$, with $M_i \in {\cal L}({\cal H}_{\rm O} \otimes {\cal H}_{\rm aux})$. The Bayesian figure of merit, the \textit{score}, is then defined as
\begin{align}\label{scoreeq2}
    {\cal S} \coloneqq \int \dd{\bm \theta} p({\bm \theta}) \sum_i p(i|{\bm \theta}) c({\bm \theta},{\bm {\hat \theta}}_i), 
\end{align}
where we took a discrete POVM for simplicity.
Note that the summation corresponds to the average cost over outcomes. The integral is then averaging over the parameter space using the prior.
By using Born's rule, one can rewrite the score as follows
\begin{align}\label{eq:scoregeneral}
    \mathcal{S} = \int \dd{\bm \theta} p({\bm \theta})  \sum_i \Tr \left[ M_i\left( \Lambda_{\bm \theta} \otimes \mathds{1} \right) \rho \right] c({\bm \theta},{\bm {\hat \theta}}_i). 
\end{align}
For this single-shot setup, solving the optimization problem amounts to finding the triplet $\{\rho^{\ast}, M^{\ast}, {\bm{\hat \theta}}^{\ast}\}$ that maximizes or minimizes the score, depending on whether $c (\boldsymbol{\theta}, \hat{\boldsymbol{\theta}}_i)$ is a reward or penalty function, respectively. Generally, a known analytical solution does not exist. However, there exists a viable algorithm, based on the formalism of higher-order operations~\cite{taranto2025higher}, to find the optimal solution numerically~\cite{PhysRevResearch.6.023305}. Inspired by this, in the present work, we extend this framework to protocols that can access several uses of the channel. This allows full exploitation of quantum correlations and joint quantum operations. Before that, let us briefly review the single-shot algorithm. 
The score in Eq.~\eqref{eq:scoregeneral} can be rewritten as 
\begin{align}\label{eq:score_tester}
    \mathcal{S} = \int \dd{\bm \theta} p({\bm \theta})  \sum_i \Tr \left[ T_i J_{\bm \theta} \right] c({\bm \theta},{\bm {\hat \theta}}_i),
\end{align}
with operator $J_{\bm \theta }\coloneqq \sum_{ij} \dyad{i}{j} \otimes \Lambda_{\bm\theta} [\dyad{i}{j}] \in \mathcal{L} \left( \mathcal{H}_\mathrm{I} \otimes \mathcal{H}_\mathrm{O} \right)$, where $\{\ket{i}\}$ is the computational basis, being the Choi-Jamiołkowski (CJ) representation of the channel $\Lambda_{\bm\theta}$. The operators $T \coloneqq  \{T_i\}_i$ with $T_i \in {\cal L}({\cal H}_{\rm I} \otimes {\cal H}_{\rm O})$, are called \textit{testers}~\cite{PhysRevLett.101.060401, PhysRevLett.101.180501, PhysRevA.77.062112, PhysRevA.80.022339}, and can be constructed from the state $\rho$ and measurement $M$ according to
\begin{align}
    T_i \coloneqq \Tr_{\rm aux} \left[(\rho^{T_{\rm aux}} \otimes \mathds{1}^{\rm O})(\mathds{1}^{\rm I} \otimes M_i^T)\right],
\end{align}
with $\rho^{T_{\rm aux}}$ being the partial transpose of $\rho$ with respect to the computational basis of the auxiliary Hilbert space. Noting that $p(i|{\bm \theta}) = \Tr (T_i J_{\bm \theta})$, one can prove Eq.~\eqref{eq:score_tester}. This expression is useful because it replaces the nonlinear dependence on $\rho$ and $M$ with a linear dependence on $T$. 
Furthermore, testers admit a simple mathematical characterization, via the following  constraints~\cite{PhysRevLett.101.060401, PhysRevLett.101.180501, PhysRevA.77.062112, PhysRevA.80.022339}:
\begin{subequations}
\begin{align}
    T_i &\geq 0, \quad \forall i\\
    \sum_i T_i &= \sigma \otimes \mathds{1}^{\rm O} \label{1copytester},
\end{align}
\end{subequations}
where $\sigma \in {\cal L}(\cal H^{\rm I})$ is a valid quantum state satisfying $\sigma \geq 0$ and $\Tr (\sigma) = 1$. Given a tester $T=\{T_i\}$, one can construct $\rho$ and $M_i$ from the optimal $T_i$ according to~\cite{PhysRevLett.101.060401, PhysRevLett.101.180501, PhysRevA.77.062112, PhysRevA.80.022339}
\begin{subequations}\label{eq:tester-realization}
\begin{align}
    \rho & \coloneqq (\mathds{1}^{\rm I}\otimes \sqrt{\sigma^T}) \sum_{ij}\ketbra{ii}{jj} (\mathds{1}^{\rm I}\otimes \sqrt{\sigma^T}),\\
    M_i & \coloneqq (\sqrt{\sigma}^{-1}\otimes \mathds{1}^{\rm O}) \ T_i^T \  (\sqrt{\sigma}^{-1}\otimes \mathds{1}^{\rm O}),~~\forall i.
\end{align}
\end{subequations}
Different sets of state and measurement yielding the same tester will also give the same score, as \eqref{eq:score_tester} just depends on the probability distribution $p (i | \boldsymbol{\theta})$. 
As a result, the optimization problem over $T$ can be solved using semidefinite programming (SDP).

\section{Multi-copy quantum parameter estimation}\label{sec:protocols}
\begin{figure*}[t]
    \centering
    \includegraphics[width=\linewidth]{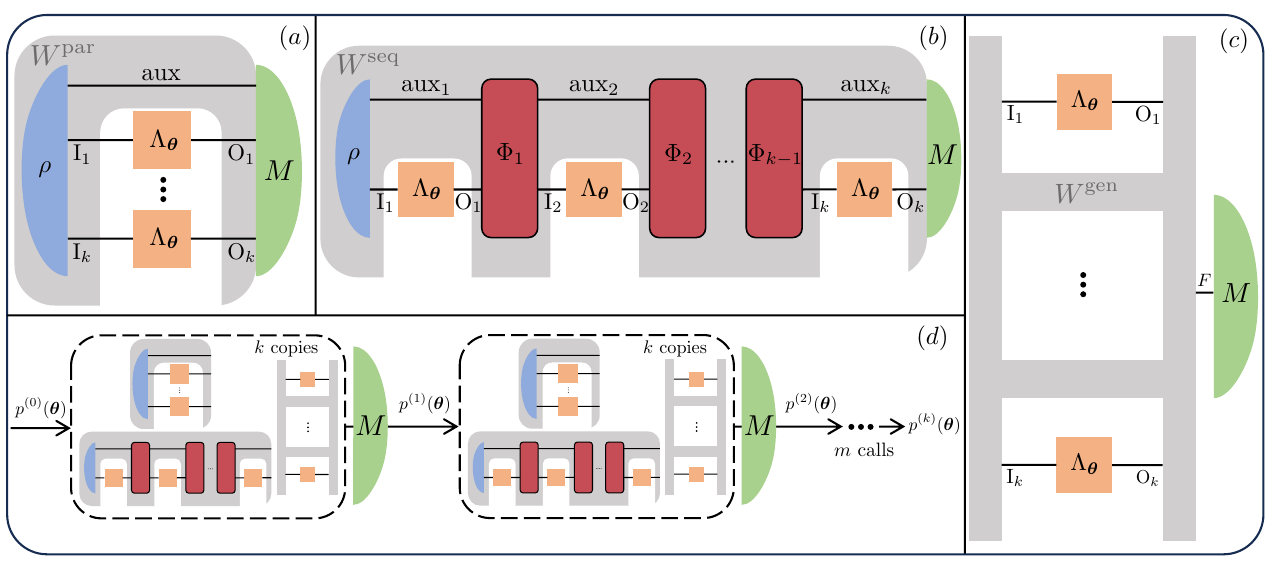}
    \caption{
    Schematic representation of the different strategies considered for Bayesian quantum parameter estimation with $k$ uses of a parameter-encoding channel $\Lambda_{\boldsymbol{\theta}}$. 
    \textit{(a) Parallel strategy} $W^{\mathrm{par}}$: a probe state $\rho$ (possibly entangled with an auxiliary system $\mathrm{aux}$) is sent through $k$ parallel copies of $\Lambda_{\boldsymbol{\theta}}$, followed by a joint measurement $M$. 
    \textit{(b) Sequential strategy} $W^{\mathrm{seq}}$: the channel uses are interleaved with intermediate operations $\Phi_1,\ldots,\Phi_{k-1}$ acting on the system and ancillary spaces, with a final measurement $M$. 
    \textit{(c) General / ICO strategy} $W^{\mathrm{gen}}$: a general higher-order operation connects the $k$ calls to $\Lambda_{\boldsymbol{\theta}}$ without a fixed causal order, before a final measurement $M$ is applied. 
    \textit{(d) Adaptive greedy strategy}: protocols are implemented in rounds with classical feedforward (no quantum memory between rounds), updating the prior and redesigning the next-round strategy based on previous measurement outcomes.}

    \label{fig:strategies}
\end{figure*}

Having reviewed the single-shot case, we are now ready to address the multi-copy scenario, the main focus of this work. 
Figure~\ref{fig:strategies} depicts the different protocols that we consider here, namely, parallel, sequential, indefinite causal order, and adaptive greedy. Generally, the idea is to use the tester formalism already discussed, but with $k$ copies~\cite{PhysRevLett.127.200504,Bavaresco2022}. This is to say, the figure of merit will read
\begin{align}\label{eq:score_multiple_copies}
    {\cal S}(T,\hat{\bm \theta}) \coloneqq   \int \dd{\bm \theta} p({\bm \theta})  \sum_i \Tr \left( T_i J_{\bm \theta}^{\otimes k} \right) c({\bm \theta},{\bm {\hat \theta}}_i),
\end{align}
where we explicitly include the dependence on the optimization arguments $T$ and $\hat{\bm \theta}$. Note that the tester elements now act on a bigger Hilbert space $T_i \in {\cal L}(({\cal H}_{\rm I}\otimes {\cal H}_{\rm O})^{\otimes k})$. 
Furthermore, the testers do not only contain the information about the inputs and measurements, they also contain all the information about the control operations performed on the probe-auxiliary systems.
The different protocols will reflect in setting the constraints on the testers, as we discuss below. We refer to the notation previously introduced in~\cite{PhysRevLett.127.200504, Bavaresco2022}.

\subsection{Parallel strategies} We first focus on parallel strategies, in which the $k$ copies of the channel are applied simultaneously on the $k$ subsystems of the state. Formally, we define this scenario with a set of parallel testers $T^\mathrm{par} \coloneqq \{ T_i^\mathrm{par} \}_{i=1}^{N_\mathrm{O}}$, with $T_i^\mathrm{par} \in {\cal L}(({\cal H}_{\rm I}\otimes {\cal H}_{\rm O})^{\otimes k})$. If we define $W^\mathrm{par} \coloneqq \sum_i T_i^\mathrm{par}$, the set of parallel testers is defined by the following constraints
\begin{subequations}\label{eq:const_par}
\begin{align}
    T_i^\mathrm{par} &\geq 0, \quad \forall i\\
    W^\mathrm{par} &= {}_{\mathrm{O}_1 \cdots \mathrm{O}_k}\!W^\mathrm{par}, \\
    \Tr W^\mathrm{par} &= d_{\mathrm{O}_1} \cdots d_{\mathrm{O}_k},
\end{align}
\end{subequations}
where each $\mathrm{O}_i$, with $i\in\{1,\ldots,k\}$, is a copy of the output system $\mathcal{H}_\mathrm{O}$, and we have defined the \textit{trace-and-replace} map $_\mathrm{X} [\cdot] \coloneqq  \Tr_X [ \cdot ] \otimes \mathds{1}_\mathrm{X} / d_\mathrm{X}$. Note that the operators $W^\mathrm{par}$ compatible with these constraints have the same form as Eq. \eqref{1copytester}, with $\sigma \in \mathcal{L} \left( \mathcal{H}_{\mathrm{I}}^{\otimes k} \right)$. Therefore, the same construction in Eqs. \eqref{eq:tester-realization} applies, yielding an explicit global probe state $\rho \in \mathcal{L} \left( \mathcal{H}_\mathrm{I}^{\otimes k} \otimes \mathcal{H}_\mathrm{aux} \right)$ and a joint POVM with elements $M_i \in \mathcal{L} \left( \mathcal{H}_\mathrm{O}^{\otimes k} \otimes \mathcal{H}_\mathrm{aux} \right)$, where generally the dimension of the auxiliary system is required to be at most $d_I^k$.

\subsection{Sequential strategies} Sequential strategies allow for adaptive uses of the channel: one can interleave the $k$ uses of $\Lambda_{\bm \theta}$ with arbitrary CPTP maps that act on the current output and an internal quantum memory, so that the input to a given use may depend on all previous outputs. As before, we can define this type of strategies with a set of sequential testers $T^\mathrm{seq} \coloneqq \{ T_i^\mathrm{seq} \}_{i=1}^{N_\mathrm{O}}$, with $T_i^\mathrm{seq} \in {\cal L}(({\cal H}_{\rm I}\otimes {\cal H}_{\rm O})^{\otimes k})$. If we define $W^\mathrm{seq} \coloneqq \sum_i T_i^\mathrm{seq}$, the set of sequential testers is defined by the following constraints
\begin{subequations}\label{eq:const_seq}
\begin{align}
    T_i^\mathrm{seq} &\geq 0, \quad \forall i \\
    \Tr W^\mathrm{seq} &= d_{\mathrm{O}_1} \cdots d_{\mathrm{O}_k}, \\
    W^\mathrm{seq} &= {}_{\mathrm{O}_k}\!W^\mathrm{seq}, \\
    {}_{\mathrm{I}_k \mathrm{O}_k} \!W^\mathrm{seq} &= {}_{\mathrm{O}_{(k-1)} \mathrm{I}_k \mathrm{O}_k}\!W^\mathrm{seq}, \\
    &\vdots \notag \\
    {}_{\mathrm{I}_2 \mathrm{O}_2 \cdots \mathrm{I}_k \mathrm{O}_k}\!W^\mathrm{seq} &= {}_{\mathrm{O}_1 \mathrm{I}_2 \mathrm{O}_2 \cdots \mathrm{I}_k \mathrm{O}_k}\!W^\mathrm{seq},
\end{align}
\end{subequations}
where just as for $\mathrm{O}_i$ previously, here $\mathrm{I}_i$, with $i\in\{1,\ldots,k\}$, is a copy of the input system $\mathcal{H}_\mathrm{I}$.
Note that this is equivalent to a $k$-slot quantum comb, since the different uses of the channel are causally ordered. In contrast to the parallel case, sequential testers do not admit, in general, a realization in terms of a single probe state and a joint POVM like in Eqs. \eqref{eq:tester-realization}. The reason is that the constraints in Eq.~\eqref{eq:const_seq} do not yield an expression like that in Eq.~\eqref{1copytester}. Hence, the corresponding explicit construction is more involved and is naturally formulated in terms of a \emph{coherent part} (encompassing state preparation and possible intermediate operations with memory) together with a final POVM (see Ref. \cite{PhysRevA.80.022339} for details).

Here, we present this construction explicitly for the case of $k=2$, which can be straightforwardly generalized to larger $k$~\cite{PhysRevA.80.022339,Ohst2026characterising}. A sequential tester $T^\mathrm{seq}=\{T_i^\mathrm{seq}\}_i$, with $T_i^\mathrm{seq}\in\mathcal{L}(\mathcal{H}_{\mathrm{I}_1}\otimes\mathcal{H}_{\mathrm{O}_1}\otimes\mathcal{H}_{\mathrm{I}_2}\otimes\mathcal{H}_{\mathrm{O}_2})$, that satisfies the constraints in Eqs.~\eqref{eq:const_seq}, can be realized by a quantum state $\rho\in\mathcal{L}(\mathcal{H}_{\mathrm{I}_1}\otimes\mathcal{H}_{\mathrm{aux}_1})$, a quantum channel in the Choi representation $\Phi\in\mathcal{L}(\mathcal{H}_{\mathrm{aux}_1}\otimes\mathcal{H}_{\mathrm{O}_1}\otimes\mathcal{H}_{\mathrm{I}_2}\otimes\mathcal{H}_{\mathrm{aux}_2})$, and a quantum measurement $M=\{M_i\}_i$, with $M_i\in\mathcal{L}(\mathcal{H}_{\mathrm{aux}_2}\otimes\mathcal{H}_{\mathrm{O}_2})$, according to
\begin{equation}
    T_i^\mathrm{seq} = \Tr_{\mathrm{aux}_1,\mathrm{aux}_2}\left[(\rho\otimes\mathds{1}_{})(\Phi^{T_{\mathrm{aux}_1,\mathrm{aux}_2}}\otimes\mathds{1})(\mathds{1}\otimes M_i^T)\right],
\end{equation}
where each identity operator is acting on the complementary spaces, with $\rho$, $\Phi$, and $M$ given by the following. Let $E=\frac{1}{d_{\rm O_2}}\tr_{{\rm O}_2}(W^\mathrm{seq})$ and $\sigma=\frac{1}{d_{\rm O_1} d_{\rm O_2}}\tr_{{\rm O}_1 {\rm I}_2{\rm O}_2}(W^\mathrm{seq})$, where again $W^\mathrm{seq} = \sum_i T_i^\mathrm{seq}$. Then,
\begin{subequations}
\begin{align}
    \rho &= (\mathds{1}^{\rm I_1}\otimes \sqrt{\sigma}^T) \sum_{ij}\ketbra{ii}{jj} (\mathds{1}^{\rm I_1}\otimes \sqrt{\sigma}^T) \\
    \Phi &= (\sqrt{\sigma}\otimes \mathds{1}^{{\rm O}_1 {\rm I}_2 {\rm aux}_2}) (\mathds{1}^{{\rm aux}_1{\rm O}_1}\otimes \sqrt{E}^T)\\ \nonumber
    & \times \sum_{ij}\ketbra{ii}{jj} (\sqrt{\sigma}\otimes \mathds{1}^{{\rm O}_1 {\rm I}_2 {\rm aux}_2}) (\mathds{1}^{{\rm aux}_1{\rm O}_1}\otimes \sqrt{E}^T) \\
    M_i &= (\sqrt{E}^{-1}\otimes \mathds{1}^{\rm O_2}) \ T_i^T \  (\sqrt{E}^{-1}\otimes \mathds{1}^{\rm O_2}),~~\forall i.
\end{align}
\end{subequations}
In this construction, $d_{\mathrm{aux}_1}=d_{\mathrm{I}}$ and $d_{\mathrm{aux}_2}=d_{\mathrm{I}}^2d_{\mathrm{O}}$. We also note that, generally, the dimension of the required auxiliary system to implement a sequential strategy grows exponentially with the number of copies $k$.

\subsection{Indefinite causal order strategies} Finally, we define general/ICO strategies as the most general transformations that map $k$ independent quantum channels
into a valid probability distribution, even when acting only on part of the input quantum channels. In terms of testers, we define $T^\mathrm{gen} \coloneqq \{ T_i^\mathrm{gen} \}_{i=1}^{N_\mathrm{O}}$, with $T_i^\mathrm{gen} \in {\cal L}(({\cal H}_{\rm I}\otimes {\cal H}_{\rm O})^{\otimes k})$. For $W^\mathrm{gen} \coloneqq \sum_i T_i^\mathrm{gen}$, the set of general testers is defined by the following constraints
\begin{subequations}\label{eq:const_gen}
\begin{align}
    T_i^\mathrm{gen} &\geq 0, \quad \forall i, \\
    \Tr \left[ W^\mathrm{gen} (J_1\otimes\ldots\otimes J_k) \right] &= 1, \label{normalizationconditionalprobs}
\end{align}
\end{subequations}
for all $J_l$, with $l\in\{1,\ldots,k\}$, that are Choi operators of quantum channels.
Note that the last condition, when $J_l=J_{\bm\theta}$ for all $l$, is equivalent to the normalization of the outcome probability distribution $p (i | {\bm\theta})$. This constraint sets no assumption on the causal order of the $k$ uses of the channel, which motivates the name of ICO strategies. For the case of $k=2$ and $k=3$ linear constraints have been derived from Eq.~\eqref{normalizationconditionalprobs} in terms of the trace-and-replace map \cite{Araujo2015Witnessing}.
The realization of a general/ICO tester can be understood in terms of a higher-order operation $W^\mathrm{gen}$, that acts on the same spaces as the tester elements $T_i^\mathrm{gen}$ and additionally on a ``future'' Hilbert space $\mathcal{H}_{\mathrm{F}}$ that succeeds all previous input and output spaces, followed by a quantum measurement $M=\{M_i\}_i$ on $\mathcal{H}_{\mathrm{F}}$~\cite{PhysRevLett.127.200504}. $W^\mathrm{gen}$ can be constructed from $T^\mathrm{gen}$ according to $W^\mathrm{gen}=\sum_i T_i^\mathrm{gen}\otimes{\ketbra{i}}$ and the measurement elements can be taken to be $M_i=\ketbra{i}$, where $\{\ket{i}\}$ is the computational basis on $\mathcal{H}_{\mathrm{F}}$, without loss of generality. As previously mentioned, while mathematically well defined, {an explicit} quantum realization of higher-order operations $W^\mathrm{gen}$ with indefinite causal order remains an open question. 
{At the same time, no logical contradictions or physical implausibilities have been found to arise from such objects, leaving open the possibility that they might be implementable and motivating the study of their potential practical consequences and advantages to metrological tasks. In addition, since these higher-order operations are mathematically the most general linear transformations that extract classical information from quantum channels, capturing greedy, parallel, and sequential strategies as particular cases, they also serve as useful tools to find upper bounds on the performance of these physically well-understood classes of strategies.}
\vspace{-0.2cm}
\subsection{Adaptive greedy strategy (hybrid quantum-classical)}\label{greedysection}
Finally, we introduce the adaptive greedy algorithm with two motivations: first, it can be seen as one of the most sophisticated protocols that does not require quantum memory, and hence can provide a good benchmark for memory-assisted protocols. Second, it can complement and boost the previous protocols in practical scenarios that one has access to limited capacity. Namely, scenarios in which, given a total number of calls to the channel $k^{\prime}=m \times k$, one has the limited capacity to perform parallel, sequential, or ICO strategies on at most $k$ copies at a time (e.g. due to memory restrictions). Yet, one can use an adaptive greedy algorithm between calls to further improve the precision via a hybrid quantum-classical approach. 

In what follows, we briefly explain this scenario for a general $k$. Numerical implementation is explained in Sec.~\ref{sec:numerical_impl}. 
Given $k'$ calls to the channel, the algorithm uses them in adaptive rounds $c \in \{ 1, \dots, m\}$ as follows:

\begin{enumerate}
    \item \textbf{Local Optimization:} 
    Using the current prior distribution $p^{(c-1)}$, calculate the optimal protocol (testers $T^{*(c)}$ and estimators $\hat{\boldsymbol{\theta}}^{*(c)}$) for the current batch of $k$ channel uses. This step is \textit{greedy} because it optimizes the immediate score for the current batch without foreseeing future calls.

    \item \textbf{Measurement:} 
    Apply the optimal protocol to the $k$ channel copies. In a practical scenario, one performs the physical measurement to obtain outcome $s$. In the numerical implementation (see Sec.~\ref{sec:numerical_impl}), this outcome is simulated probabilistically based on the sample's ``true'' parameter(s).

    \item \textbf{Bayesian Update:} 
    Use the obtained outcome $s$ to compute the posterior distribution via Bayes' rule (Eq. \eqref{bayesrule}). This posterior becomes the prior $p^{(c)}$ for the next adaptive round.

    \item \textbf{Repeat:} 
    Iterate steps 1--3 until the total budget of $k' = m \times k$ calls is exhausted.
\end{enumerate}
\section{Numerical implementation}\label{sec:numerical_impl}
As already mentioned, expressing the score in terms of testers allows for an efficient optimization via an SDP. In principle, the optimal tester has at most $d_{\max} = (d_{\rm I}^k \times d_{\rm O}^k)^2$ many outcomes, since the optimal state and POVM should be extremal, and since extremal POVMs in dimension $d$ have at most $d^2$ many outcomes. Nonetheless, it is sometimes beneficial to inflate the number of outcomes such that $N_\mathrm{O} \geq d_{\rm max}$, which is a useful approach in finding the optimal estimators as well~\cite{PhysRevResearch.6.023305}. 

Furthermore, the integral over the hypothesis parameters $\boldsymbol{\theta}$ in Eq.~\eqref{eq:score_multiple_copies} can be approximated by a finite sum. To this aim, we discretize the parameter space by replacing the continuous domain $\Theta$ with a finite set of sample points $\{\boldsymbol{\theta}_j\}_{j=1}^{N_{\rm H}}$, where $\boldsymbol{\theta}_j \in \Theta$ for all $j$. The choice of this set depends on the nature of the prior distribution $p(\boldsymbol{\theta})$ and the cost function. For distributions close to uniform, a regular mesh (grid) is often suitable. However, for sharper priors---which naturally arise as posteriors after several measurement updates---more adaptive strategies are beneficial. Namely, importance sampling (Monte Carlo) can be used to allocate samples according to the probability density~\cite{press2007numerical}.

Regardless of the sampling method, this discretization allows us to enumerate the $q$-dimensional vector of parameters using a single linear index $j\in\{1, \dots, N_H\}$. This unifies the notation, allowing us to treat single and multiparameter estimation problems identically by using a single summation. Furthermore, the estimators naturally form a finite list, as they are in a one-to-one correspondence with the POVM outcomes, so we define $\{\hat{\boldsymbol{\theta}}_i\}_{i=1}^{N_{\rm O}}\eqqcolon {\hat {\bm \theta}}$. Consequently, the continuous score in Eq.~\eqref{eq:score_multiple_copies} is approximated by the discrete sum 
\begin{equation}
    \widetilde{\cal S}(T, {\hat {\bm \theta}}) \simeq \sum_{j=1}^{N_{\rm H}} \sum_{i=1}^{N_{\rm O}} p_j c({\bm\theta}_j, \hat{{\bm\theta}}_i) \text{Tr}\left( T_i J_{{\bm \theta}_j}^{\otimes k} \right),
    \label{eq:score_approximate}
\end{equation}
where we defined $p_j \coloneqq p({\bm \theta} = {\bm \theta_j})/\sum_{j=1}^{N_{\rm H}}p({\bm \theta} = {\bm \theta_j})$. The approximation error in this sum typically scales as $\mathcal{O} (N_{\rm H}^{-2/q})$ for a regular mesh and $\mathcal{O} (N_{\rm H}^{-1/2})$ for random (importance) sampling \cite{Caflisch1998}.

Equation~\eqref{eq:score_approximate} is a good approximation to the exact score, as it can be made very close to the true score by simply increasing $N_\mathrm{H}$, which is computationally not demanding. 
Our optimization problem is thus reduced to

\begin{figure}[t!]
\begin{myalgobox}{\small Finding the optimal Bayesian protocol}{box:seesaw}
    \small
    \begin{enumerate}[leftmargin=*]
        \item \textbf{Initialization:} 
        Select an initial set of estimators denoted by ${\hat {\bm \theta}}^{(0)}$. Set the iteration counter $c=1$.

        \item \textbf{Step 1 (Optimize Testers):} 
        Fix the estimators to their current value ${\hat {\bm \theta}}^{(c-1)}$. Solve the SDP problem to find the optimal testers $T^{(c)}$ that optimize the score:
        \begin{equation*}
             T^{(c)} = \underset{T}{\text{arg opt}} \ {\widetilde {\cal S}}(T, {\hat {\bm \theta}}^{(c-1)})
        \end{equation*}
        subject to constraints given by Eqs.~\eqref{eq:const_par}, \eqref{eq:const_seq}, or \eqref{eq:const_gen}.

        \item \textbf{Step 2 (Optimize Estimators):} 
        Fix the testers to the optimal values $T^{(c)}$ found in the previous step. Optimize the estimators to find ${\hat {\bm \theta}}^{(c)}$:
        \begin{equation*}
            {\hat {\bm \theta}}^{(c)} = \underset{\hat{\theta}}{\text{arg opt}} \ {\widetilde {\cal S}}(T^{(c)}, {\hat {\bm \theta}}).
        \end{equation*}

        \item \textbf{Convergence Check:} 
        Evaluate the change in the score:
        \begin{equation*}
            \Delta = |{\widetilde {\cal S}}(T^{(c)}, {\hat {\bm \theta}}^{(c)}) - {\widetilde {\cal S}}(T^{(c-1)}, {\hat {\bm \theta}}^{(c-1)})|
        \end{equation*}
        If $\Delta < \epsilon$ (where $\epsilon$ is a pre-defined tolerance), terminate and output $T^{*} = T^{(c)}$, ${\hat {\bm \theta}}^* = {\hat {\bm \theta}}^{(c)}$, and ${\widetilde{\cal S}^*} = {\widetilde{\cal S}^*}(T^{(c)},{\hat {\bm \theta}}^{(c)})$. Otherwise, increment $c \leftarrow c+1$ and return to Step 1.
    \end{enumerate}
\end{myalgobox}
\end{figure}

\begin{align*}
    \widetilde {\cal S}^*
    \coloneqq &\operatorname*{opt}_{\{T,{\hat {\bm \theta}}\}}\ \widetilde {\cal S}(T,{\hat {\bm \theta}}) \\
    &\text{ s.t. } \{T,{\hat {\bm \theta}}\}  \text{ satisfy Eqs. } \eqref{eq:const_par},\ \eqref{eq:const_seq},\ \text{or } \eqref{eq:const_gen},
    \label{eq:opt-problem}
\end{align*}
where ${\rm opt}$ denotes either maximization or minimization depending on the cost function.

\begin{figure}[t!]
\begin{myalgobox}{\small The adaptive greedy algorithm}{box:greedy}
    \small
    \begin{enumerate}[leftmargin=*]
        \item \textbf{Initialization:} Define cost $c(\boldsymbol{\theta}, \hat{\boldsymbol{\theta}})$ and the initial prior $\{p_j\}_{j=1}^{N_\mathrm{H}}$. 
        \item \textbf{Initial Optimization (independent of the Monte Carlo round)}: Find optimal testers $T^{*(1)}$ and estimators ${\hat {\bm \theta}}^{*(1)}$ using $\{p_j\}_{j=1}^{N_\mathrm{H}}$ according to Box~\ref{box:seesaw}.
        \item \textbf{Monte Carlo simulation Loop:} Set the total number of Monte Carlo trajectories, $N_m \gg 1$. For $l \in\{ 1, \dots, N_m\}$:
        \begin{enumerate}[leftmargin=1.5em, label=(\alph*)]
            \item \textit{Initialize the prior:} (Re)set the current prior to $\{p^{(0)}_j\}_{j=1}^{N_\mathrm{H}} = \{p_j\}_{j=1}^{N_\mathrm{H}}$.
            \item \textit{Sample True Parameter:} Pick $j_{\text{true}}$ from $\{p_j^{(0)}\}_{j=1}^{N_\mathrm{H}}$. Set
            $ \boldsymbol{\theta}^{(l)}_{\text{true}} = {\bm\theta}_{j_{\text{true}}} $.

            \item \textit{Adaptive Rounds:} For copy $c \in\{ 1, \dots, m\}$:
            \begin{enumerate}[leftmargin=1.2em, label=\roman*.]
                \item \textbf{Optimization:} If $c$>1, find optimal testers $T^{*(c)}$ and estimators ${\hat {\bm \theta}}^{*(c)}$ using prior $\{p^{(c-1)}_j\}_{j=1}^{N_\mathrm{H}}$ according to Box~\ref{box:seesaw}.
                
                \item \textbf{Simulation:} Calculate outcome probabilities:
                \begin{equation*}
                   P(\nu | \boldsymbol{\theta}^{(l)}_{\text{true}}) = \text{Tr}\big( T_\nu^{(c)} J_{\boldsymbol{\theta}^{(l)}_{\text{true}}}^{\otimes k} \big).
                \end{equation*}
                Randomly select outcome label $s$ based on $P(s | \boldsymbol{\theta}^{(l)}_{\text{true}})$.
                
                \item \textbf{Update Estimator:} Set $\hat{\boldsymbol{\theta}}^{(l)} = \hat{\bm \theta}_{s}^{*(c)}$.
                
                \item \textbf{Update Score:} $\widetilde{\cal S}^{(l)} = c(\boldsymbol{\theta}^{(l)}_{\text{true}}, \hat{\boldsymbol{\theta}}^{(l)})$.
                
                \item \textbf{Bayesian Update:} Update prior. For $j\in\{1, \dots, N_\mathrm{H}\}$:
                \begin{equation*}
                    p^{(c)}_j = \frac{p^{(c-1)}_j\text{Tr}\big( T_s^{(c)} J_{{\bm \theta}_j}^{\otimes k} \big)}{\mathcal{N}}
                \end{equation*}
                where $\mathcal{N}$ is the normalization factor.
                
                \item 
                {\bf Optional for large $m$:} Resample (particle filtering).
            \end{enumerate}
        \end{enumerate}
    
        \item \textbf{Final Calculation:} Average the score:
        \begin{equation}\label{eq:average_MC_greedy_score}
            \bar{\cal S} = \frac{1}{N_m} \sum_{l=1}^{N_m} \widetilde{\cal S}^{(l)}.
        \end{equation}
    \end{enumerate}
\end{myalgobox}
\end{figure}

For a fixed set of estimators,  we can formulate the problem of finding the optimal testers as an SDP, which is highly efficient. By taking $N_\mathrm{O} \gg d_{\max}$ and choosing uniformly picked fix estimators, one can also assure that the problem is fully solved without requiring to optimize over estimators~\cite{PhysRevResearch.6.023305}. However, we take the seesaw approach as explained in Box~\ref{box:seesaw}, which is numerically more efficient.

All codes implementing this algorithm for the three protocols are freely available in our code repository~\cite{github}.

Lastly, let us remark on the numerical implementation of the adaptive greedy protocol. In this case, we take a Monte Carlo approach to simulate the optimal score, which is explained in Box~\ref{box:greedy}.

By increasing the number of Monte Carlo simulations $N_m$, the average score will converge to the optimal greedy score $\bar{S} \overset{N_m \gg 1}{\longrightarrow} {\widetilde S}^*_{\rm greedy}$. 
If in step (c)-i one only optimizes over the estimators, while keeping the testers fixed, we call the protocol \textit{non-adaptive}, which we will use in some examples below as a benchmark to specifically demonstrate the usefulness of the adaptive greedy approach.

\section{Examples}\label{sec:examples}
\begin{table*}[t!]
\centering
{
\caption{Summary of the case studies considered in Sec.~\ref{sec:examples}. For each estimation problem, we list the unknown parameters, the prior distribution, the Bayesian cost or reward function, and the hierarchy observed among the different strategy classes.}
\label{tab:casestudies}
\begin{tabularx}{\textwidth}{lXXX}
\toprule\toprule
 & SU(2) estimation & Thermometry & Noisy SU(2) estimation \\
\midrule
Parameters
& $\boldsymbol{\theta} = (\theta_x,\theta_y,\theta_z)$
& $\theta$
& $\boldsymbol{\theta} = (\theta_x,\theta_y,\theta_z)$
\\
\midrule
Parameter encoding
& $U_{\bm\theta} \coloneqq e^{-i \boldsymbol{\theta} \cdot \boldsymbol{\sigma}}$
& Eq.~\eqref{MEthermometry}
& $U_{\bm\theta} \coloneqq e^{-i \boldsymbol{\theta} \cdot \boldsymbol{\sigma}}$ + Eq.~\eqref{krausAD}
\\
\midrule
Prior
& Haar-uniform over SU(2)
& Uniform over $\theta/\epsilon \in [1,20]$
& Haar-uniform over SU(2)
\\
\midrule
Cost/reward
& $\frac{1}{4}\Tr\!\left( J_{\boldsymbol{\theta}} J_{\hat{\boldsymbol{\theta}}_i} \right)$
& $\left( \frac{\theta - \hat{\theta}_i}{\theta} \right)^2$
& $\frac{1}{4}\Tr\!\left( J^U_{\boldsymbol{\theta}} J^U_{\hat{\boldsymbol{\theta}}_i} \right)$
\\
\midrule
Performance hierarchy
& Greedy $<$ PAR $=$ SEQ $=$ ICO
& Greedy $=$ PAR $=$ SEQ $=$ ICO
& Greedy $<$ PAR $<$ SEQ $<$ ICO
\\
\bottomrule\bottomrule
\end{tabularx}}
\end{table*}
\subsection{SU(2) multiparameter estimation}\label{sec:examples_su2}
We will now use our methods to solve several problems. Some of these have analytical solutions, namely the SU(2) multiparameter estimation problem, and are hence ideal for benchmarking our methods. In thermometry, we find no hierarchy between strategies, with the greedy approach at $k=1$ matching sequential protocols. In noisy SU(2), in contrast, we find a strict hierarchy across the considered protocols. We leave a fourth example on quantum phase estimation to Appendix~\ref{app:phase_est}, where we also observe a strict hierarchy between the different strategies. Furthermore, our methods are not bound to specific prior distributions. We remind that even if one starts with an elegant prior that enables partial analytical solutions (e.g., in the SU(2) problem), in a greedy strategy posteriors do not necessarily respect any symmetries. Our methods treat the problem equally, regardless of the prior. 

{To ease the readability of the section, in Table \ref{tab:casestudies} we summarize the results we obtained for the three case studies, namely, the cost function and performance hierarchy between strategies for each example.} 

Let us start with a multiparameter estimation problem. To this aim, take a (single-qubit) unitary transformation $U_{\bm\theta} \in \mathrm{SU (2)}$ to be parametrized by a vector of unknown angles, $\boldsymbol{\theta} \coloneqq (\theta_x, \theta_y, \theta_z)$ such that $U_{\bm\theta} \coloneqq e^{-i \boldsymbol{\theta} \cdot \boldsymbol{\sigma}}$, with $\boldsymbol{\sigma} \coloneqq (\sigma_x, \sigma_y, \sigma_z)$ being the vector of Pauli matrices. This estimation task is crucial, for instance, in spatial orientation and reference-frame alignment~\cite{RevModPhys.79.555, PhysRevA.70.030301}. In the context of quantum metrology, several applications include two-mode interferometry~\cite{PhysRevA.91.032103} and Ramsey-type spectroscopy~\cite{PRXQuantum.4.020333, Schulte2020ramsey}.

The channel corresponding to the unitary that encodes the parameters is then given by $\Lambda_{\bm\theta} [\cdot] \coloneqq U_{\bm\theta} \cdot U_{\bm\theta}^\dagger$, with $d_{\rm I} = d_{\rm O} = 2$.
Note that the SU(2) group is isomorphic to the group of quaternions with norm 1. That is, for any $U \in {\rm SU(2)}$ there exists a vector $q \coloneqq (q_0, q_1, q_2, q_3) \in \mathbb{R}^{4}$ with $ \norm{q}^2 = q^T q = 1$ such that $U \equiv q_0 I - i(q_1 \sigma_x + q_2 \sigma_y + q_3 \sigma_3)$. Conversely, any $q \in \mathbb{R}^4$ with $q^T q = 1$ defines an element of $\mathrm{SU (2)}$ through the same formula.

Regarding the cost function, it is conventional to take the fidelity between the Choi operators of the true parameter and its estimator~\cite{RAGINSKY200111, PhysRevA.71.062310}. Namely,
\begin{equation}\label{fidelityreward}
    c ({\bm\theta}, \hat{{\bm\theta}}_i) = \dfrac{1}{4} \Tr (J_{\bm\theta} J_{\hat{{\bm\theta}}_i}) = \dfrac{1}{4} \abs{\Tr \left( U^{\dagger}_{\bm\theta} U_{\hat{\bm\theta}_i} \right)}^2 = \left( q_{\bm \theta}^T q_{\hat{\bm\theta}_i} \right)^2 
\end{equation}
Note that $0 \leq c ({\bm\theta}, \hat{{\bm\theta}}_i) \leq 1$, and we are dealing with a \textit{maximization} problem. In a memory-assisted scenario with $k$ simultaneous calls to the channel, the optimal tester has $d_{\rm max} = 4^{2k}$ many outcomes.
As we discuss in Appendix \ref{optimalestimatorsSU2appendix}, one can conveniently identify the optimal estimator for a given observation by its quaternion representation,
\begin{equation}\label{estimatoroptimizationsu2}
    q_{\hat{\bm \theta}_i}^* \coloneqq \underset{q}{\rm arg max} ~q^T K q,
\end{equation}
with $K \coloneqq \int \dd{\bm \theta} p({\bm \theta}|i) q_{\bm \theta} q_{\bm \theta}^T$, being $q_{\bm \theta}$ the quaternion associated to $U_{\bm \theta}$. Note that this problem reduces to finding the eigenstate of $K$ with the maximum eigenvalue. As $K$ is a $4 \times 4$ matrix, the optimal estimators can be found efficiently with Eq.~\eqref{estimatoroptimizationsu2}.

Furthermore, in the specific case that the prior over SU(2) is Haar random~\cite{Spengler_2012}, the optimal score can be found analytically and is achieved by parallel strategies. In particular~\cite{PhysRevA.69.050303, Quintino2022deterministic},  
\begin{align}\label{eq:SU2_exact}
    {\cal S}^* = \cos^2 \left( \dfrac{\pi}{k + 3} \right).
\end{align}
The optimal parallel strategy achieving this score was established earlier in Ref.~\cite{PhysRevA.64.050302}. Furthermore, neither sequential~\cite{PhysRevLett.101.180501} nor ICO~\cite{Bavaresco2022,Hayashi2025indefinitecausal} strategies can outperform the parallel case.

\begin{figure} 
    \centering
    \includegraphics[width=\linewidth]{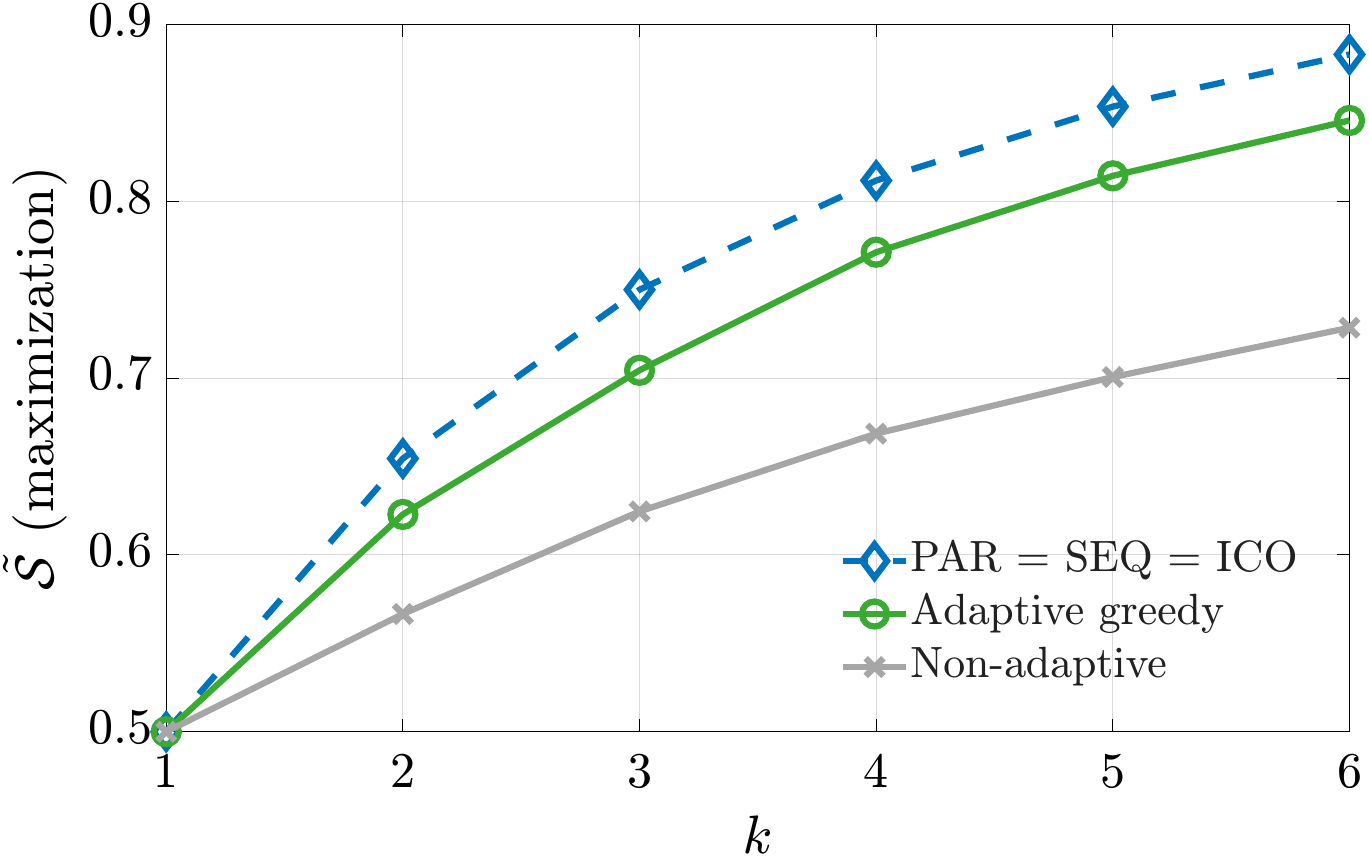}
    \caption{
    Comparison between different strategies in the SU(2) multiparameter estimation.  The optimal protocol (blue dashed) outperforms the greedy strategy (solid green)  which in turn outperforms the non-adaptive and non-memory-assisted strategy (gray solid). 
    The results suggest that the adaptive greedy strategy with $k=5$ calls is as good as the best quantum memory-assisted strategy with $k=4$ calls.
    The adaptive greedy and the non-adaptive scores are obtained using $N_m = 10^4$ random Monte Carlo simulations, with  $N_\mathrm{H} = 8000$ and $N_\mathrm{O} = 27$.}
    \label{fig:su2_seq_vs_adaptive}
\end{figure}

We apply our methods to determine the optimal score for this problem using the Haar random prior distribution, details of which are provided in Appendix~\ref{Haarappendix}. This serves as a robust benchmark due to the existence of an analytical solution for this prior. Taking $k=2$, we find that  ${\cal {\widetilde S}}^* \approx 0.6545  = \cos^2(\pi/5)$---matching the exact solution~\eqref{eq:SU2_exact}---for parallel, sequential, and indefinite causal order strategies. 
We then benchmark the performance of the adaptive greedy algorithm against the optimal strategy in Figure~\ref{fig:su2_seq_vs_adaptive}. The performance gap between the two strategies indicates that global operations are indeed advantageous, however, at the limit of few copies the greedy algorithm remains a viable alternative; e.g., to get the same score as the one that the best quantum strategy achieves with $k=4$ simultaneous calls to the channel, one can use only $k=5$ single calls with the adaptive greedy strategy. To complete the analysis, we simulate a non-adaptive strategy without memory in between the uses of the channels. In this approach, we optimize the testers for the first channel use and fix them for all remaining consecutive calls. This strategy generally under-performs compared to the greedy strategy due to the lack of adaptivity, as it is also seen in the figure.

\subsection{Thermometry}\label{sec:examples_sthermometry}
In this example we focus on quantum thermometry~\cite{Mehboudi2019,DePasquale2018}. Contrary to our previous example which involved a unitary encoding, the channel that encodes the temperature is inherently dissipative. 
In this setting, the unknown parameter is the temperature $\theta$ of a sample (or thermal reservoir) at equilibrium. We take the channel's input and output dimensions to be $d_{\rm I} = d_{\rm O} = 2$, i.e., a two-level system (qubit). While initially at time $t=0$ the input to the channel can be correlated to auxiliary systems, and other probes entering copies of the channel, the probes and the sample are assumed to be uncorrelated. After a fixed interaction time $t$, a joint measurement on the probes and auxiliaries is performed in order to estimate the bath temperature. In order to identify the encoding channel, recall that in a single call case the input to the channel---i.e., the marginal input of the probe-auxiliary $\rho^{p}_\theta \coloneqq \Tr_A(\rho_\theta)$---evolves according to a standard Markovian quantum master equation~\cite{Breuer2007} 
\begin{equation}\label{MEthermometry}
    \dot{\rho}_\theta^p (t) = -i \comm{H}{\rho_\theta^p (t)} + \Gamma_\mathrm{in} \mathcal{D} \left[ \sigma_+ \right] \rho_\theta^p (t) + \Gamma_\mathrm{out} \mathcal{D} \left[ \sigma_- \right] \rho_\theta^p (t),
\end{equation}
where $H = \epsilon \dyad{1}$ is the Hamiltonian of the probe, $\sigma_-$ ($\sigma_+$) is the lowering (raising) Pauli operator, and $\mathcal{D}[A]\rho = A\rho A^\dagger - \frac{1}{2} \acomm{A^\dagger A}{\rho}$ denotes the usual Lindblad dissipator, describing the coupling to the bath. The only temperature-dependent quantities are the excitation and relaxation rates $\Gamma_{\mathrm{in}}$ and $\Gamma_{\mathrm{out}}$, which for a bosonic reservoir are given by $\Gamma_{\mathrm{in}} = J(\epsilon) N_{\rm B} (\epsilon)$ and $\Gamma_{\mathrm{out}} = J(\epsilon) \left(1 + N_{\rm B} (\epsilon) \right)$, where $J(\epsilon)$ is the bath spectral density and $N_{\rm B} (\epsilon) = (e^{\epsilon/\theta}-1)^{-1}$ is the thermal occupation number of the bosonic mode at energy $\epsilon$. We note that our methods can be seamlessly applied to fermionic baths as well. We also note that the setup covers equilibrium thermometry ($t\to \infty$) where the initial state and control operations are expected not to matter, only the measurement choice to have an impact, as well as non-equilibrium scenarios where in principle all parts of the protocol may be determinant.

For this problem, the channel $\Lambda_\theta (t)$ is generated by the dynamics given in Eq.~\eqref{MEthermometry}, and the analytical expressions for both the action of the channel on an initial state and the Choi operator have already been derived in the literature \cite{PhysRevResearch.6.023305} (see also Appendix \ref{appendixthermometry} for the explicit expressions). Regarding the cost function in thermometry, it is conventional to take either the mean square logarithmic error or the relative mean square error~\cite{PhysRevLett.127.190402,PhysRevLett.128.130502,PhysRevA.105.042601,PhysRevA.104.052214}. Here, we take the latter, given by
\begin{equation}
    c (\theta, \hat{\theta}_i) = \left( \dfrac{\theta - \hat{\theta}_i}{\theta} \right)^2,
\end{equation}
which needs to be \textit{minimized} to find the optimal strategy. We expect a similar behavior for the mean square logarithmic error. An advantage of this figure of merit is that the optimal measurements are PVMs and thus $d_{\rm max} = 4^k$. Furthermore, conveniently, the optimal estimator is known to be~\cite{PhysRevA.104.052214}
\begin{equation}
    \hat{\theta}_i^* = \dfrac{\ev{1/\theta_j}^{(i)}}{\ev{1/\theta_j^2}^{(i)}},
\end{equation}
 where $\ev{\cdot}^{(i)}$ indicates taking the mean over the posterior probability distribution $p(\theta|i)$. 

\begin{figure}[t!]
    \centering
    \includegraphics[width=\linewidth]{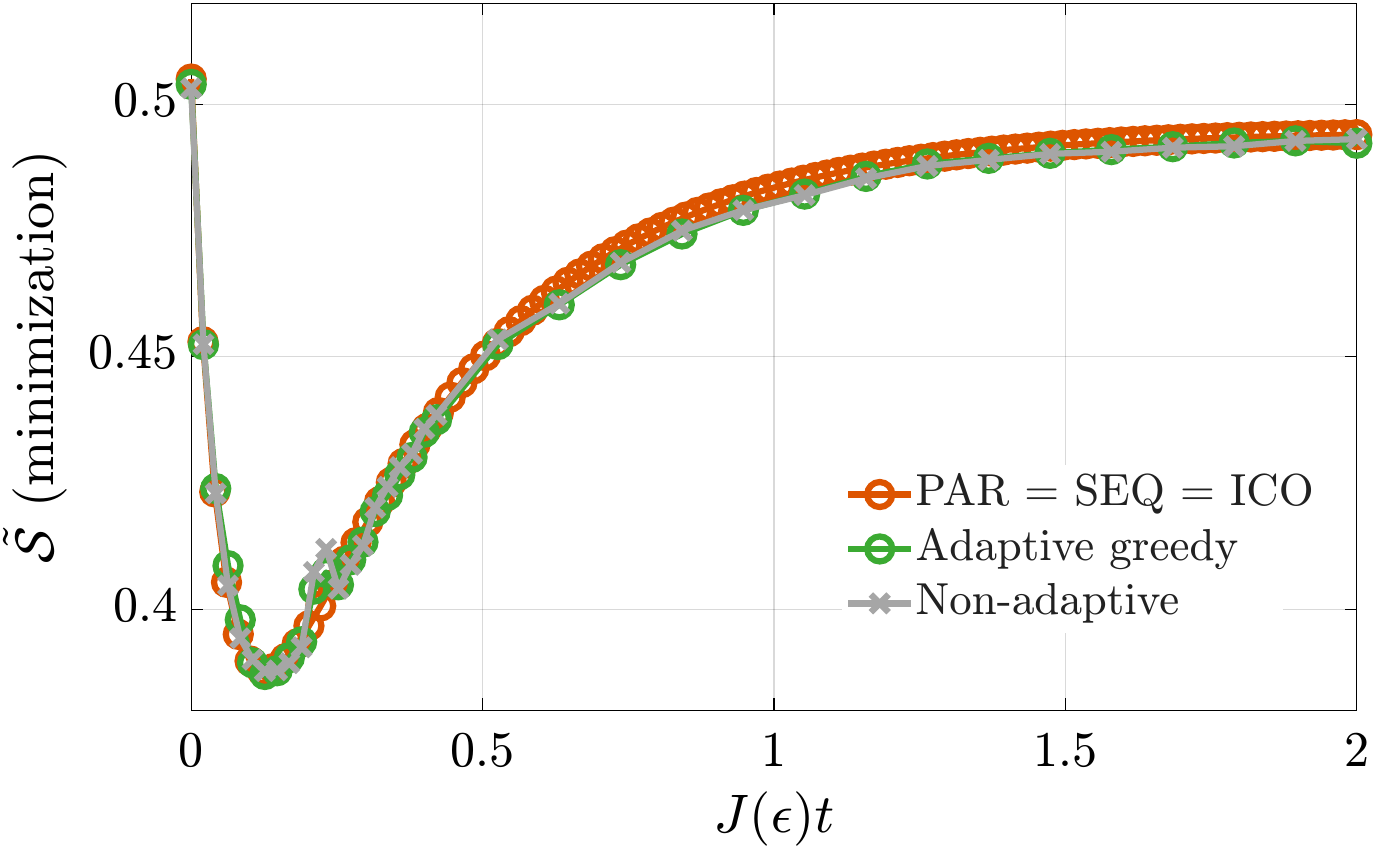}
    \caption{Minimum approximate score $\widetilde{\mathcal{S}}$ in a thermometry estimation task for a greedy algorithm (green solid line) compared to a sequential strategy (red solid line) for $k=2$ copies of the channel. For the optimization we have taken $N_\mathrm{H} = 2500$ and $N_\mathrm{O} = 20 ~(N_O=4)$ outputs for the quantum memory-assisted (adaptive greedy/non-adaptive) strategies, respectively. For the adaptive greedy/non-adaptive strategy, the simulation has been run for $N_m = 10^6$ Monte Carlo iterations. The prior is given by a uniform distribution in the range $\theta/\epsilon \in [1, 20]$.}
    \label{fig:thermometrygreedy}
\end{figure}

In Figure~\ref{fig:thermometrygreedy}, we benchmark the optimal {thermometry} score for parallel, sequential, and ICO strategies against the adaptive greedy strategy. We observe that the latter equals the general sequential case for all times, which indicates that the strategy is optimal without the need for entanglement between the different uses of the channel (note that this does not rule out entanglement with auxiliary systems, which can give an advantage over separable states in the transient regime \cite{PhysRevResearch.6.023305}). Moreover, even non-adaptive classical strategies perform equally good, hence adaptive manipulation of the setting seems unnecessary. 
Let us further point out that the parallel approach is indeed expected to scale with $k$ in the asymptotic limit. To this aim, one can study the QFI---a relevant figure of merit in the asymptotic limit of many channel calls $k \gg 1$. As we show in Appendix \ref{appendixthermometry}, the QFI for the parallel strategies scales at most linearly with the number of calls to the channel, therefore not beating the shot-noise limit and being realizable with a classical strategy. {Lastly, although we assumed a uniform prior, we observed a similar behavior (no quantum advantage) for several other priors (data not shown). We therefore guess that this observation about thermometry via the channel in Eq.~\eqref{MEthermometry} is general and extends to any quadratic cost function, given that quadratic cost functions are equivalent (up to the prior distribution) under reparameterization.}

\subsection{Dissipative encoding of SU(2)}\label{sec:examples_dissipative_su2}
As our last example we take a more complex channel, in which we showcase a strict hierarchy between the memory-assisted strategies. Namely, that the sequential strategy outperforms the parallel strategy, while being outperformed by an ICO strategy. In this case, the greedy algorithm underperforms compared to all the memory-assisted strategies, as it happened in the noiseless unitary estimation problem considered before. The parameters to be estimated and the figure of merit are the same ones as the SU(2) example. 

Following Ref.~\cite{PhysRevLett.130.070803}, we consider a family of channels obtained by first encoding a vector of desired parameters $\boldsymbol{\theta}$ via the SU(2) unitary $U_{\boldsymbol{\theta}}$ as described previously, followed by an undesired amplitude-damping (AD) channel that adds noise to the system. We note that, unlike the thermometry problem, this noise here is not inherent to the encoding; it rather arises from imperfect transmission. The Kraus operators of the amplitude damping channel are given by~\cite{Nielsen2012}
\begin{equation}\label{krausAD}
    {\bar K}_1 = \begin{pmatrix}
        1 & 0 \\
        0 & \sqrt{1-p}
    \end{pmatrix}, \quad 
    {\bar K}_2 = \begin{pmatrix}
        0 & \sqrt{p} \\
        0 & 0
    \end{pmatrix},
\end{equation}
where $p\in[0,1]$ quantifies the decay ($p=0$ corresponds to the identity map, while $p=1$ maps every input state to the ground state, hence wiping out the encoded information). The overall channel is therefore $
\Lambda_{\boldsymbol{\theta}} = \mathrm{AD} \circ \mathcal{U}_{\boldsymbol{\theta}}$
which admits Kraus operators $\{K_{1} = {\bar K}_1U_{\boldsymbol{\theta}},\,K_{2} = {\bar K}_2U_{\boldsymbol{\theta}}\}$.

\begin{figure}
    \centering
    \includegraphics[width=\linewidth]{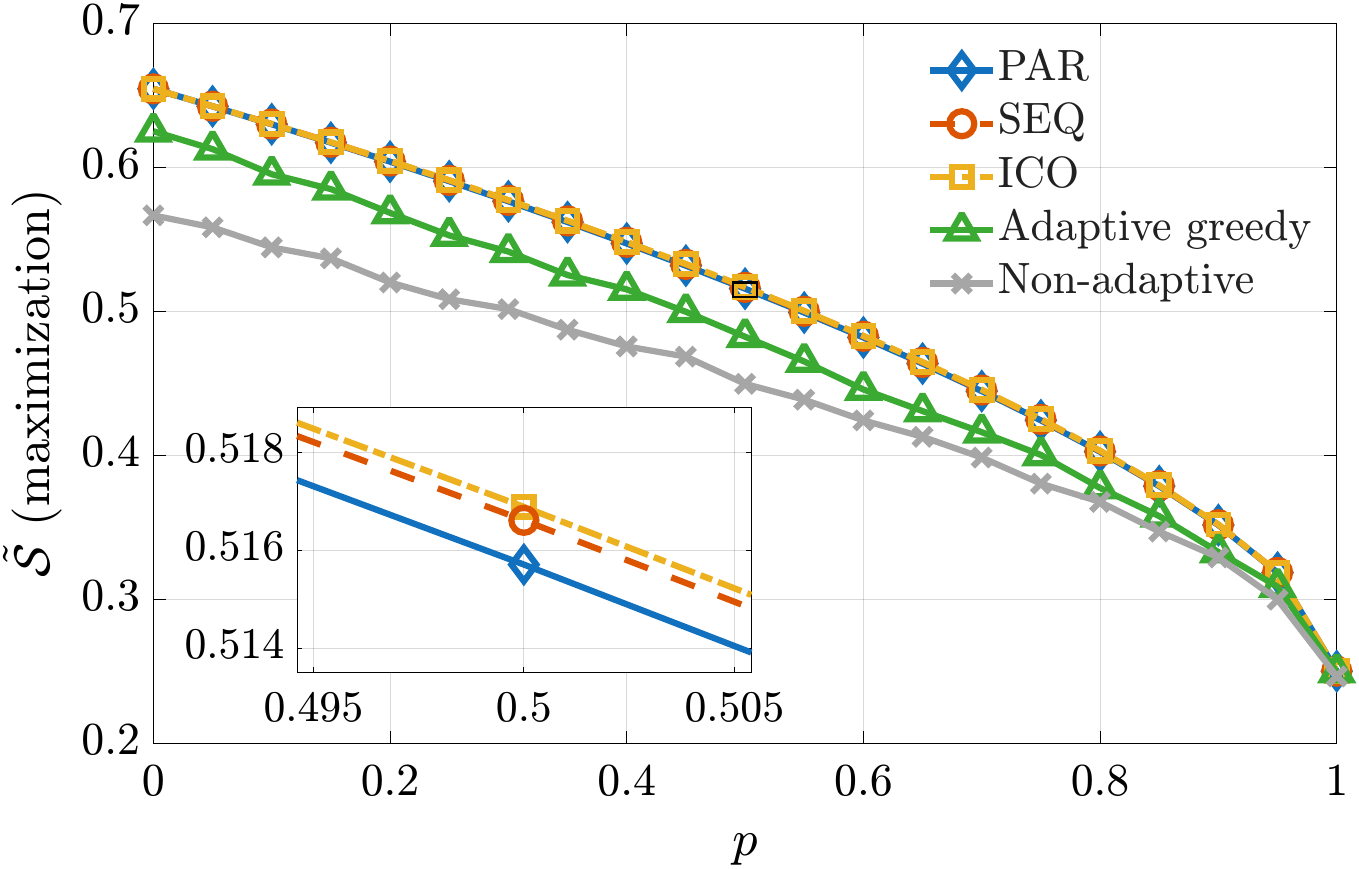}
    \caption{
    Maximum approximate score $\widetilde{\mathcal{S}}$ in a SU(2) estimation task followed by an AD channel as a function of the damping parameter $p$ for $k = 2$ copies of the channel. We zoom in around $p=0.5$ to showcase the gaps between strategies. The simulation has been run for $N_m = 10^4$ Monte Carlo iterations, $N_\mathrm{H} = 8000$ and $N_\mathrm{O} = 1000 ~(N_O=27)$ outputs for the quantum memory-assisted (adaptive greedy/non-adaptive) strategies.}
    \label{fig:hierarchysu2}
\end{figure}

As already advanced, our goal is to estimate the set of parameters $\boldsymbol{\theta}$ encoded in the unitary $U_{\boldsymbol{\theta}}$ (and not to estimate the full noisy channel).
The cost function that we consider remains the same as that given by Eq.~\eqref{fidelityreward}, i.e., $c({\bm \theta},{\hat{\bm \theta}}_i) = \frac{1}{4} {\rm Tr}(J^U_{\bm \theta}J^U_{\hat{\bm \theta}_i})$, where to avoid confusion, we denoted $J_{\bm \theta}^U$ to be the Choi-Jamiolkowski representation of the unitary channel $U_{\bm \theta}$. In contrast, $J_{\bm \theta}^p$ is the Choi-Jamiolkowski representation of the total channel $\Lambda_{\bm \theta}$, which is what we need to substitute for calculation of the outcome probabilities---that is ${\rm Tr}[T_i (J^p_{\bm \theta})^{\otimes k}]$ and ${\rm Tr}[T_i (J^p_{{\bm \theta}_j})^{\otimes k}]$ in Eqs.~\eqref{eq:score_multiple_copies} and~\eqref{eq:score_approximate}, respectively.
In Figure \ref{fig:hierarchysu2} we plot the maximum approximate score versus $p$. For the extreme cases of $p \in \{0,1\}$, one observes no hierarchy between parallel, sequential and ICO strategies, while greedy underperforms all three. Namely, for $p=0$ the encoding is ideal, and the problem reduces to that of Sec.~\ref{sec:examples_su2}. Furthermore, at $p=1$, all collected data is erased, and therefore the choice of the strategy does not matter after all. In fact, the only contribution to the score comes from the prior information, which as we show in Appendix \ref{approximatescorep1} is analytically equal to $1/4$. Notably, for intermediate values of $p$, a strict hierarchy PAR<SEQ<ICO is observed, despite the score values being very close.

\section{Discussion}\label{sec:discussion}
In this work, we have developed a comprehensive framework for Bayesian quantum parameter estimation in the regime of finite data. By exploiting the formalism of quantum testers and higher-order operations, we provided a versatile algorithm and the accompanying numerical tools capable of approximating the optimal metrological strategy---encompassing the probe state, the control operations, the POVM elements, and the estimators---for arbitrary channel uses. This approach allowed us to compare the performance of parallel, sequential, and indefinite causal order strategies, as well as an adaptive greedy scheme based on classical feedforward.

Our results reveal a rich landscape of performance hierarchies. We demonstrated that the superiority of a particular class of strategies is highly problem-dependent. For the ideal SU(2) encoding, the hierarchy collapses, rendering parallel strategies as effective as the most general ICO protocols. Here we show that adaptive greedy strategies, which rely on classical instead of quantum memory, provide a great alternative for such strategies in this problem: while a $k$-copy adaptive greedy strategy performs worse than a $k$-copy parallel or sequential strategy, we observe that greedy strategies using $k+1$ copies are already able to nearly match or in some cases even outperform the best parallel strategy with $k$ uses. This has been explicitly observed in the regime of $k\in\{2,\ldots,6\}$.

Conversely, in the presence of noise, such as the SU(2) channel followed by amplitude damping, a strict hierarchy emerges in the $2$-copy case ($ \text{PAR} < \text{SEQ} < \text{ICO} $). Interestingly, for dissipative thermometry, we found that the adaptive greedy strategy, and even a non-adaptive strategy performs on par with the optimal sequential strategy. This result is of significant practical value, as it suggests that for certain dissipative tasks, classical feedforward suffices to attain the ultimate precision limits without requiring long-lived quantum memories to maintain entanglement across channel uses. Identifying the crucial feature of the encoding process (the quantum channel) that leads to a hierarchy between the different strategies deserves further exploration.

A distinct advantage of our Bayesian framework lies in its operational feasibility compared to frequentist approaches. Frequentist bounds, such as the Quantum Cram\'er-Rao Bound, often rely on the QFI~\cite{Toth_2014,doi:10.1142/S0219749909004839}. While powerful, the QFI formalism can suffer from the incompatibility of optimal measurements. Specifically, in the multiparameter setting, the optimal input, measurements, and estimators for different parameters may suffer incompatibility issues~\cite{Belliardo_2021,PhysRevA.105.062442,hu2024control}. Consequently, the bound may not be attainable by any physical strategy. In contrast, our approach optimizes the Bayesian score directly, rather than a bound on it. By construction, the resulting strategy yields a specific initial state, a specific set of intermediate control operations, a single valid POVM, and an explicit set of estimators. 

Finally, we address the scalability of numerical methods in quantum metrology. The semidefinite programming approach employed here provides arbitrary precision but suffers from the curse of dimensionality. The size of the testers scales exponentially with the number of channel uses $k$, limiting the direct application of SDPs to small $k$. While in this work we limited to $k=2$, we believe with current state of the art computational resources, one could solve the problem for $k\leq4$ as well, for qubit-qubit channels, but not far beyond. To tackle the regime of higher channel uses, integrating tensor network techniques into the Bayesian framework remains as an interesting future direction. These techniques have been successfully applied to calculate the QFI for large systems by efficiently representing the relevant quantum states and channels~\cite{Kurdzialek_2025}, as well as to the problem of channel discrimination~\cite{sieniawski2026adaptive}. { Adapting them to the Bayesian formalism is a natural next step, one that could also assist in Bayesian estimation with many-body probes.}

\subsection*{Acknowledgments}
We are thankful to Marcus Huber for his insights on Haar random unitaries and their characterization.
We are also thankful to Simon Milz and Marco Túlio Quintino for publicly sharing their numerical codes on the higher-order operations formalism~\cite{milz2024characterising,githubMTQ}. 
The authors acknowledge TU Wien Bibliothek for financial support through its Open Access Funding Programme.
This research was funded in part by the Austrian Science Fund (FWF) [“NOQUS,” Grant DOI: 10.55776/PAT1969224].
{
\section*{DATA AVAILABILITY}
All the necessary codes, parameters, and instructions to reproduce the 
simulations in our work can be found in this repository \cite{zenodo}.}
\bibliography{biblio} 

\onecolumngrid
\appendix
\color{black}{}
\section{Details on the SU(2) estimation problem}\label{app:SU2_example}
\subsection{The optimal estimators}\label{optimalestimatorsSU2appendix}
We devote this section to derive the analytical expression for the optimal estimators in the SU(2) problem (namely, Eq.~\eqref{estimatoroptimizationsu2}). Recalling Eq.~\eqref{scoreeq2} and using the reward function for the SU(2) estimation problem (already in the quaternion representation) we have
\begin{equation}
    \mathcal{S} = \sum_{i=1}^{N_\mathrm{O}} \int \dd{\boldsymbol{\theta}} p (\boldsymbol{\theta}) p (i | \boldsymbol{\theta}) \left( q_{\boldsymbol{\theta}}^T q_{\hat{\boldsymbol{\theta}}_i} \right)^2 = \sum_{i=1}^{N_\mathrm{O}} \int \dd{\boldsymbol{\theta}} p (\boldsymbol{\theta}) p (i | \boldsymbol{\theta}) q_{\hat{\boldsymbol{\theta}}_i}^T q_{\boldsymbol{\theta}} q_{\boldsymbol{\theta}}^T q_{\hat{\boldsymbol{\theta}}_i} = \sum_{i=1}^{N_\mathrm{O}} q_{\hat{\boldsymbol{\theta}}_i}^T \left[ \int \dd{\boldsymbol{\theta}} p (\boldsymbol{\theta}) p (i | \boldsymbol{\theta}) q_{\boldsymbol{\theta}} q_{\boldsymbol{\theta}}^T \right] q_{\hat{\boldsymbol{\theta}}_i}.
 \end{equation}
 Therefore, for a fixed set of testers---hence a fixed set of $\{p(i|\bm \theta)\}_i$---the optimal estimators can be found  by maximizing the above $\mathcal{S}$ over the set of $q_{\hat{\boldsymbol{\theta}}_i}$:
 \begin{equation}
     q_{\hat{\bm \theta}_i^*}\coloneqq \underset{q_{\hat{\boldsymbol{\theta}}_i}}{\rm arg~max} ~~q_{\hat{\boldsymbol{\theta}}_i}^T \left[ \int \dd{\bm \theta} p({\bm \theta}|i) q_{\bm \theta} q_{\bm \theta}^T \right] q_{\hat{\boldsymbol{\theta}}_i}.
 \end{equation}
If we define
 \begin{equation}
     K \coloneqq  \int \dd{\bm \theta} p({\bm \theta}|i) q_{\bm \theta} q_{\bm \theta}^T,
 \end{equation}
 we recover exactly Eq.~\eqref{estimatoroptimizationsu2}.
\subsection{Further details on the Haar measure}\label{Haarappendix}
In the SU(2) example, we took the Haar random prior distribution. In our parametrization, we defined $U_{\boldsymbol{\theta}} = e^{-i \boldsymbol{\theta} \cdot \boldsymbol{\sigma}}$, which is a conventional notation. To benchmark our numerical results with analytical expressions, we need to express the Haar measure in this notation, $\dd{U_{\bm \theta}} = \dd{\bm \theta}p(\bm \theta)$. 
To this aim, we note that a Haar measure is already available in the literature in a different parametrization. Namely, Ref. \cite{Spengler_2012}, represents any unitary in SU(2) as
\begin{equation}\label{marcusparametrization}
    U_{\bm \lambda} = e^{i \sigma_z \lambda_3} e^{i \sigma_y \lambda_2} e^{i \sigma_z \lambda_1},
\end{equation}
with the Haar measure given by
\begin{equation}\label{oldhaar}
    \dd{U}_{\bm \lambda} = \dfrac{1}{\pi^2} \sin \lambda_2 \cos \lambda_2 \dd{\lambda_1} \dd{\lambda_2} \dd{\lambda_3}.
\end{equation}
To proceed further, we need to find the change of variables that relates both representation of the SU(2) unitaries. Let us start by explicitly expanding Eq.~\eqref{marcusparametrization} using the identity $e^{i \lambda \sigma_i} = \cos \lambda \mathds{1} + i \sin \lambda \sigma_i$:
\begin{align}
    U_{\bm \lambda} & = \left( \cos \lambda_3 \mathds{1} + i \sin \lambda_3 \sigma_z \right) \left( \cos \lambda_2 \mathds{1} + i \sin \lambda_2 \sigma_y \right) \left( \cos \lambda_1 \mathds{1} + i \sin \lambda_1 \sigma_z \right) = c_0 \mathds{1} + i (c_x \sigma_x + c_y \sigma_y + c_z \sigma_z),
\end{align}
with
\begin{align}
    c_0 &= \cos (\lambda_2) \cos (\lambda_1 + \lambda_3), \\
    c_x &= \sin (\lambda_2) \sin (\lambda_3 - \lambda_1), \\
    c_y &= \sin (\lambda_2) \cos (\lambda_3 - \lambda_1), \\
    c_z &= \cos (\lambda_2) \sin (\lambda_1 + \lambda_3).
\end{align}
Furthermore, we note that we can similarly write $U_{\bm \theta} = \cos \left( \norm{\boldsymbol{\theta}} \right) \mathds{1} - i \sin \left( \norm{\boldsymbol{\theta}} \right) \boldsymbol{n}_\theta \cdot \boldsymbol{\sigma}$. Comparing the coefficients of the different Pauli operators and the identity operator, we get
\begin{align}
    \cos \left( \norm{\boldsymbol{\theta}} \right) &= \cos (\lambda_2) \cos (\lambda_1 + \lambda_3) \label{constraintA10} \\
    \theta_x &= - \dfrac{\norm{\boldsymbol{\theta}}}{\sin \left( \norm{\boldsymbol{\theta}} \right)} \sin (\lambda_2) \sin (\lambda_3 - \lambda_1) \\
    \theta_y &= - \dfrac{\norm{\boldsymbol{\theta}}}{\sin \left( \norm{\boldsymbol{\theta}} \right)} \sin (\lambda_2) \cos (\lambda_3 - \lambda_1) \\
    \theta_z &= - \dfrac{\norm{\boldsymbol{\theta}}}{\sin \left( \norm{\boldsymbol{\theta}} \right)} \cos (\lambda_2) \sin (\lambda_1 + \lambda_3).
\end{align}
To express the measure in Eq.~\eqref{oldhaar} in terms of the new parameters, we just need to compute the Jacobian of the transformation $\{ \lambda_1, \lambda_2, \lambda_3 \} \mapsto \{ \theta_x, \theta_y, \theta_z \}$. To do that, we first need to calculate the following partial derivatives:
\begin{align}
    \pdv{\theta_x}{\lambda_1} &= \sin (\lambda_2) \left[ \dfrac{r}{\sin (r)} \cos (\lambda_3 - \lambda_1) - \dfrac{\cos (\lambda_2) \sin (\lambda_1 + \lambda_3) (\sin (r) - r \cos (r))}{\sin^3 (r)} \sin (\lambda_3 - \lambda_1) \right] \\
    \pdv{\theta_x}{\lambda_2} &= -\sin (\lambda_3 - \lambda_1) \left[ \dfrac{r}{\sin (r)} \cos (\lambda_2) + \dfrac{\sin (\lambda_2) \cos (\lambda_1 + \lambda_3) (\sin (r) - r \cos (r))}{\sin^3 (r)} \sin (\lambda_2) \right] \\
    \pdv{\theta_x}{\lambda_3} &= -\sin (\lambda_2) \left[ \dfrac{r}{\sin (r)} \cos (\lambda_3 - \lambda_1) + \dfrac{\cos (\lambda_2) \sin (\lambda_1 + \lambda_3) (\sin (r) - r \cos (r))}{\sin^3 (r)} \sin (\lambda_3 - \lambda_1) \right] \\
    \pdv{\theta_y}{\lambda_1} &= -\sin (\lambda_2) \left[ \dfrac{r}{\sin (r)} \sin (\lambda_3 - \lambda_1) + \dfrac{\cos (\lambda_2) \sin (\lambda_1 + \lambda_3) (\sin (r) - r \cos (r))}{\sin^3 (r)} \cos (\lambda_3 - \lambda_1) \right] \\
    \pdv{\theta_y}{\lambda_2} &= -\cos (\lambda_3 - \lambda_1) \left[ \dfrac{r}{\sin (r)} \cos (\lambda_2) + \dfrac{\sin (\lambda_2) \cos (\lambda_1 + \lambda_3) (\sin (r) - r \cos (r))}{\sin^3 (r)} \sin (\lambda_2) \right] \\
    \pdv{\theta_y}{\lambda_3} &= \sin (\lambda_2) \left[ \dfrac{r}{\sin (r)} \sin (\lambda_3 - \lambda_1) - \dfrac{\cos (\lambda_2) \sin (\lambda_1 + \lambda_3) (\sin (r) - r \cos (r))}{\sin^3 (r)} \cos (\lambda_3 - \lambda_1) \right] \\
    \pdv{\theta_z}{\lambda_1} &= -\cos (\lambda_2) \left[ \dfrac{r}{\sin (r)} \cos (\lambda_3 + \lambda_1) + \dfrac{\cos (\lambda_2) \sin (\lambda_1 + \lambda_3) (\sin (r) - r \cos (r))}{\sin^3 (r)} \sin (\lambda_3 + \lambda_1) \right] = \pdv{\theta_z}{\lambda_3} \\
    \pdv{\theta_z}{\lambda_2} &= -\sin (\lambda_3 + \lambda_1) \left[ \dfrac{\sin (\lambda_2) \cos (\lambda_1 + \lambda_3) (\sin (r) - r \cos (r))}{\sin^3 (r)} \cos (\lambda_2) - \dfrac{r}{\sin (r)} \sin (\lambda_2) \right],
\end{align}
with $r \coloneqq  \norm{\boldsymbol{\theta}}$. Now, the determinant of the Jacobian matrix can be shown to be
\begin{align}
    \abs{\det \pdv{(\theta_x, \theta_y, \theta_z)}{(\lambda_1, \lambda_2, \lambda_3)}} &= \dfrac{1}{32} r^{2} \csc^{3} r (
  64 r \cos\lambda_{1}\cos^{2}\lambda_{2}\cos\lambda_{3}\sin\lambda_{2}
  + 32 \cos(2\lambda_{1})\cos^{3}\lambda_{2}\cos(2\lambda_{3})(-1 + r \cot r)\csc r\sin\lambda_{2}\nonumber
  \\ 
  &- 4(-1 + r \cot r)\csc r(6\sin(2\lambda_{2}) - \sin(4\lambda_{2}))
  - 64 r \cos^{2}\lambda_{2}\sin\lambda_{1}\sin\lambda_{2}\sin\lambda_{3}
  \nonumber\\
  &- 32 \cos^{3}\lambda_{2}(-1 + r \cot r)\csc r\sin(2\lambda_{1})\sin\lambda_{2}\sin(2\lambda_{3})
),
\end{align}
which can be simplified using the constraint in Eq.~\eqref{constraintA10} to
\begin{equation}
    \abs{\det \pdv{(\theta_x, \theta_y, \theta_z)}{(\lambda_1, \lambda_2, \lambda_3)}} = 2 \left( \dfrac{r}{\sin (r)} \right)^2 \sin (\lambda_2) \cos (\lambda_2).
\end{equation}
Therefore, Eq.~\eqref{oldhaar} can be transformed into
\begin{align}
    \dd{U}_{\bm \lambda} &= \dfrac{1}{\pi^2} \sin \lambda_2 \cos \lambda_2 \dd{\lambda_1} \dd{\lambda_2} \dd{\lambda_3} = \dfrac{1}{\pi^2} \sin \lambda_2 (\theta_x, \theta_y, \theta_z) \cos \lambda_2 (\theta_x, \theta_y, \theta_z) \abs{\det \pdv{(\theta_x, \theta_y, \theta_z)}{(\lambda_1, \lambda_2, \lambda_3)}} \dd{\theta_x} \dd{\theta_y} \dd{\theta_z} \nonumber \\
    &= \dfrac{1}{2 \pi^2} \left( \dfrac{\sin r}{r} \right)^2 \dd{\theta_x} \dd{\theta_y} \dd{\theta_z} \equiv \dd U_{\bm \theta}. \label{newhaar}
\end{align}
Therefore, if we identify $\dd {\bm \theta} \equiv \dd\theta_x\dd\theta_y\dd\theta_z$, then $p({\bm \theta}) = \frac{1}{2\pi^2} [(\sin r)/r]^2$.
One can confirm that Eq. \eqref{newhaar} is properly normalized
\begin{equation}
    \int \dd{U}_{\bm \theta} = \dfrac{1}{2 \pi^2} \int_0^\pi \dd{r} r^2 \left( \dfrac{\sin r}{r} \right)^2 \int \dd{\Omega} = \dfrac{1}{2 \pi^2} \cdot \dfrac{\pi}{2} \cdot 4 \pi = 1.
\end{equation}
Therefore, if we do not want to overcount points (i.e. that the Haar measure is normalized), the vector of parameters must be inside the ball generated by $(\theta^x)^2 + (\theta^y)^2 + (\theta^z)^2 < \pi$.

\section{The Thermometry Problem}\label{appendixthermometry}
Let us first write explicitly the CJ operator of the thermometry channel~\cite{PhysRevResearch.6.023305}
\begin{equation}\label{CJthermometry}
    J_\theta = \begin{pmatrix}
        \frac{1 + N_{\rm B} (\epsilon) (e^{-(\Gamma_\mathrm{in} + \Gamma_\mathrm{out}) t} + 1)}{2 N_{\rm B} (\epsilon) + 1} & 0 & 0 & e^{-\frac{\Gamma_\mathrm{in} + \Gamma_\mathrm{out}}{2} t} \\
        0 & \frac{(N_{\rm B} (\epsilon) + 1) (1 - e^{-(\Gamma_\mathrm{in} + \Gamma_\mathrm{out}) t})}{2 N_{\rm B} (\epsilon) + 1} & 0 & 0 \\
        0 & 0 & \frac{N_{\rm B} (\epsilon) (1 - e^{-(\Gamma_\mathrm{in} + \Gamma_\mathrm{out}) t})}{2 N_{\rm B} (\epsilon) + 1} & 0 \\
        e^{-\frac{\Gamma_\mathrm{in} + \Gamma_\mathrm{out}}{2} t} & 0 & 0 & \frac{N_{\rm B} (\epsilon) + e^{-(\Gamma_\mathrm{in} + \Gamma_\mathrm{out}) t} (1 + N_{\rm B} (\epsilon))}{2 N_{\rm B} (\epsilon) + 1}
    \end{pmatrix}
\end{equation}
which can be written in terms of Kraus operators $\{ K_\alpha \}_\alpha$ as follows:
\begin{gather}\label{eq:thermometry_kraus}
    J_\theta = \sum_{\alpha=1}^4 \Vect (K_\alpha) \Vect (K_\alpha)^\dagger,\\
    K_1 = \sqrt{p} \begin{pmatrix} 1 & 0 \\ 0 & \sqrt{\gamma} \end{pmatrix},~~ 
    K_2 = \sqrt{p} \begin{pmatrix} 0 & \sqrt{1-\gamma} \\ 0 & 0 \end{pmatrix}, \\
    K_3 = \sqrt{1-p} \begin{pmatrix} \sqrt{\gamma} & 0 \\ 0 & 1 \end{pmatrix}, ~~
    K_4 = \sqrt{1-p} \begin{pmatrix} 0 & 0 \\ \sqrt{1-\gamma} & 0 \end{pmatrix},
\end{gather}
with the vectorization operator defined as $\Vect (M) \coloneqq  \sum_{ij} M_{ij} \ket{i,j}$ for a matrix $M = \sum_{ij} M_{ij} \dyad{i}{j}$, and with $p \coloneqq  (N_{\rm B} (\epsilon) + 1)/(2 N_{\rm B} (\epsilon) + 1)$ and $\gamma \coloneqq  e^{-(\Gamma_\mathrm{in} + \Gamma_\mathrm{out}) t}$.
With this at hand, we can study whether there is an advantage or not in using \textit{parallel} quantum strategies for the thermometry problem presented in Sec. \ref{sec:examples_sthermometry}. 
Following \cite{DemkowiczDobrzanski2012Elusive}, we note that the QFI when using $k$ parallel calls to the channel is bounded as
\begin{equation}
    \mathcal{F}^{(k)} \le 4 \min_{\tilde{K}} \left\{ k \norm{\alpha_{\tilde{K}}} + k(k-1) \norm{\beta_{\tilde{K}}}^2 \right\},
\end{equation}
with $\norm{ \bullet}$ being the operator norm of $\bullet$, the minimization running over all possible  Kraus representations of the channel ${\widetilde K} = \{{\widetilde K_i}\}_i$. Furthermore, $\alpha_{\tilde{K}} \coloneqq \sum_i \dot{\tilde{K}}_i^\dagger \dot{\tilde{K}}_i$ and $\beta_{\tilde{K}} \coloneqq i \sum_i \dot{\tilde{K}}_i^\dagger \tilde{K}_i$, where the dot represents the derivative with respect to the parameter being estimated.
We note that only the second term can be proportional to $k^2$, which enables super-linear (Heisenberg) scaling.
In our case, since the channel parameters depend directly on $N_{\rm B}$, the problem of estimating $\theta$ is equivalent to estimating $N_{\rm B}$ up to a change of variable; we can relate the QFI for the two parameters via the chain rule: $\mathcal{F}^{(k)}_T = \mathcal{F}^{(k)}_{N_{\rm B}} \left( \frac{dN_{\rm B}}{dT} \right)^2$. For the purpose of analyzing the QFI scaling with the number of probes $k$, it is sufficient to analyze the properties of the channel with respect to $N_{\rm B}$.

To summarize, a sufficient condition to rule out super-linear scaling is to find any valid Kraus representation $\tilde{K}$ for which the operator norm vanishes, i.e., $\norm{\beta_{\tilde{K}}} = 0$. If we can show this for our canonical choice of Kraus operators in Eq.~\eqref{eq:thermometry_kraus}, then the Heisenberg term vanishes, and the QFI is fundamentally limited to linear scaling, $\mathcal{F}^{(k)}_{N_B} \propto k$.

Now, let us show that the operator $\beta = i \sum_{k=1}^4 \dot{K}_k^\dagger K_k$ is the zero matrix for our thermometry problem. First, we calculate the derivative of each Kraus operator with respect to $N$:
\begin{align*}
    \dot{K}_1 &= \frac{\dot{p}}{2\sqrt{p}} \begin{pmatrix} 1 & 0 \\ 0 & \sqrt{\gamma} \end{pmatrix} + \sqrt{p} \begin{pmatrix} 0 & 0 \\ 0 & \frac{\dot{\gamma}}{2\sqrt{\gamma}} \end{pmatrix}, \\
    \dot{K}_2 &= \frac{\dot{p}}{2\sqrt{p}} \begin{pmatrix} 0 & \sqrt{1-\gamma} \\ 0 & 0 \end{pmatrix} + \sqrt{p} \begin{pmatrix} 0 & \frac{-\dot{\gamma}}{2\sqrt{1-\gamma}} \\ 0 & 0 \end{pmatrix}, \\
    \dot{K}_3 &= \frac{-\dot{p}}{2\sqrt{1-p}} \begin{pmatrix} \sqrt{\gamma} & 0 \\ 0 & 1 \end{pmatrix} + \sqrt{1-p} \begin{pmatrix} \frac{\dot{\gamma}}{2\sqrt{\gamma}} & 0 \\ 0 & 0 \end{pmatrix}, \\
    \dot{K}_4 &= \frac{-\dot{p}}{2\sqrt{1-p}} \begin{pmatrix} 0 & 0 \\ \sqrt{1-\gamma} & 0 \end{pmatrix} + \sqrt{1-p} \begin{pmatrix} 0 & 0 \\ \frac{-\dot{\gamma}}{2\sqrt{1-\gamma}} & 0 \end{pmatrix}.
\end{align*}
Plugging these results, it is easy to see that $\beta = i \sum_k \dot{K}_k^\dagger K_k = \mathbf{0}$, with $\mathbf{0}$ being the null operator, whose norm is trivially zero. This proves that the term responsible for super-linear scaling of the QFI vanishes, and that the precision of thermometry using this channel is therefore fundamentally limited to the standard quantum limit, scaling at best linearly with the number of probes. The same scaling is achievable with strategies that require no memory. While this behaviour should not necessarily hold in a Bayesian formalism, at the limit of large $k$, we expect the QFI to become a relevant figure of merit and the linear scaling to hold. Our results in the main text show that the classical adaptive greedy strategy is equally good even for a small number of calls, i.e., $k=2$.
\section{Details of the dissipative SU(2) encoding}\label{approximatescorep1}
In the dissipative SU(2) example, the score for the two extremal points $p=0$ and $p=1$ can be solved analytically. They can help us monitor whether our algorithm is being implemented properly in that example. While $p=0$ should trivially give the optimal score in Eq. \eqref{eq:SU2_exact}, as the amplitude damping contribution is just the identity channel, the $p=1$ case should be studied in more detail. For a general state $\rho$, the action of the channel at $p=1$ reads
\begin{equation}
    \Lambda_\mathrm{\bm \theta}^{(p=1)} \left[ \rho \right] = \sum_i {\bar K}_i^{(p=1)} U_{\bm \theta}\rho U_{\bm \theta}^{\dagger} \left({\bar K}_i^{(p=1)}\right)^\dagger.
\end{equation}
Applying the Kraus operators given in Eq. \eqref{krausAD} we get:
\begin{equation}\label{eq:CJ_p1}
    \Lambda_\mathrm{\bm \theta}^{(p=1)} \left[ \rho \right] = \dyad{0} \rho \dyad{0} + \dyad{0}{1} \rho \dyad{1}{0} = \dyad{0},
\end{equation}
as $\Tr (\rho) = 1$. Therefore, the channel is independent of ${\bm \theta}$ and all information on $\bm \theta$ is erased as expected. We recall that the score is given by
\begin{equation}
    \mathcal{S}^{(p=1)} = \sum_{i=1}^{N_\mathrm{O}} \int \dd{\bm \theta} p({\bm \theta}) c( {\bm \theta}, \hat{{\bm \theta}}_i) \Tr \left( J^{(p=1)}_{\bm \theta} T_i \right).
\end{equation}
We note that $\Tr \left( J^{(p=1)}_{\bm \theta} T_i \right) = \Tr \left( J^{(p=1)}_* T_i \right)\eqqcolon q_i$, with $J_*$ the Choi-Jamiołkowski representation of the channel in Eq.~\eqref{eq:CJ_p1} being independent of ${\bm \theta}$ as we showed above. We also note that $\sum_i q_i = 1$. Then, we can write the score as
\begin{equation}
    \mathcal{S}^{(p=1)} = \dfrac{1}{d^2} \sum_{i=1}^{N_\mathrm{O}} q_i \Tr \left[ \left( \int \dd{\bm \theta} p({\bm \theta}) J^{U}_{\bm \theta} \right) J^{U}_{\hat{\bm \theta}_i} \right].
\end{equation}
Now, by using that (see below for a proof)
\begin{equation}\label{choiaverage}
    \int \dd{\bm \theta} p ({\bm \theta}) J^U_{\bm \theta} = \dfrac{\mathds{1}}{d},
\end{equation}
the score will be
\begin{equation}
    \mathcal{S}^{(p=1)} = \dfrac{1}{d^3} \sum_{i=1}^{N_\mathrm{O}} q_i \Tr ( J^U_{\hat{\bm \theta}_i} ) = \dfrac{1}{d^2} \sum_{i=1}^{N_\mathrm{O}} q_i = \dfrac{1}{d^2},
\end{equation}
as we wanted to show. It is left to prove that Eq.~\eqref{choiaverage} holds. By using $J^U_{\boldsymbol{\theta}} = \sum_{ij} \dyad{i}{j} \otimes U_{\bm \theta} \dyad{i}{j} U_{\bm \theta}^\dagger$ we have,
\begin{equation}
    \int \dd{\bm \theta} p({\bm \theta}) J^U_{\bm \theta} = \int \dd{\bm \theta} p({\bm \theta}) \left( \sum_{ij} \dyad{i}{j} \otimes U_{\bm \theta} \dyad{i}{j} U_{\bm \theta}^\dagger \right) = \sum_{ij} \left( \dyad{i}{j} \otimes \int \dd{\bm \theta} p({\bm \theta}) U_{\bm \theta} \dyad{i}{j} U_{\bm \theta}^\dagger \right).
\end{equation}
Now, writing the unitary in the computational basis as $U_{\bm \theta} = \sum_{ab} [U_{\bm \theta}]_{ab} \dyad{a}{b}$ we get:
\begin{align}
    \int \dd{\bm \theta} p({\bm \theta}) J^U_{\bm \theta} &= \sum_{ij} \left[ \dyad{i}{j} \otimes \int \dd{\bm \theta} p({\bm \theta}) \left( \sum_{ab} [U_{\bm \theta}]_{ab} \dyad{a}{b} \right) \dyad{i}{j} \left( \sum_{a'b'} [U^*_{\bm \theta}]_{a'b'} \dyad{b'}{a'} \right) \right] \nonumber\\
    &=\sum_{ij} \left[ \dyad{i}{j} \otimes \int \dd{\bm \theta} p({\bm \theta}) \left( \sum_{ab} [U_{\bm \theta}]_{ab} \delta_{bi} \ket{a} \right) \left( \sum_{a'b'} [U^*_{\bm \theta}]_{a'b'} \delta_{j b'} \bra{a'} \right) \right] \nonumber\\
    &= \sum_{ij} \left( \dyad{i}{j} \otimes \int \dd{\bm \theta} p({\bm \theta}) \sum_{a a'} [U_{\bm \theta}]_{ai} [U^*_{\bm \theta}]_{a'j} \dyad{a}{a'} \right) \nonumber\\
    &= \sum_{ij} \left( \dyad{i}{j} \otimes \sum_{a a'} \dyad{a}{a'} \int \dd{\bm \theta} p({\bm \theta}) [U_{\bm \theta}]_{ai} [U^*_{\bm \theta}]_{a'j} \right) \nonumber\\
    &= \dfrac{1}{d} \sum_{ij} \left( \dyad{i}{j} \otimes \sum_{a a'} \dyad{a}{a'} \delta_{a a'} \delta_{ij} \right) = \dfrac{\mathds{1}}{d},
\end{align}
where we used that
\begin{equation}
    \int \dd{\bm \theta} p({\bm \theta}) [U_{\bm \theta}]_{ai} [U^*_{\bm \theta}]_{a'j} = \dfrac{1}{d} \delta_{a a'} \delta_{i j},
\end{equation}
which is the Haar orthogonality relation for the matrix elements of a unitary matrix (see e.g., theorem 5.8 of~\cite{Folland1995AbstractHarmonicAnalysis} or corollary 2.4 of~\cite{collins_integration_2006}).

\section{Dissipative phase estimation}\label{app:phase_est}
In this example, we refer to the problem presented in Ref.~\cite{PhysRevLett.130.070803} which is treated within a \textit{frequentist} approach. There, a hierarchy in the optimal QFI is observed. The parameter is a phase that is encoded by a unitary followed by noise. Namely,
\begin{equation}
    U_{\theta} = e^{-i \theta t \sigma_z/2},
\end{equation}
followed by the amplitude damping channel with Kraus operators that we defined earlier in Eq. \eqref{krausAD}. 
For the phase estimation problem, the cost function is conventionally taken as
\begin{equation}\label{cosreward}
    c (\theta, \hat{\theta}_i) = \cos^2 \left( \dfrac{\theta - \hat{\theta}_i}{2} \right).
\end{equation}

In Ref.~\cite{PhysRevLett.130.070803}, a hierarchy between the different quantum-assisted strategies is reported when $0<p<1$.
Here, we rather treat this problem within a \textit{Bayesian} approach and establish some connections with those in the frequentist case in the next section. 

To this aim, following Ref.~\cite{e20090628} we take the following prior:
\begin{equation}\label{customprior}
    p (\theta) = \dfrac{ \exp \left[ \alpha \sin^{2} \left( \pi \tfrac{\theta - \theta_{\min}}{\theta_{\max} - \theta_{\min}} \right) \right] - 1}
     { (\theta_{\max} - \theta_{\min}) \left[ e^{\alpha/2} I_{0}\!\left(\tfrac{\alpha}{2}\right) - 1 \right] },
\end{equation}
with $I_0$ being the $0$-th order modified Bessel function of the first kind. Here, $\theta_{\min} = 0$ and $\theta_{\max} = 2 \pi$. Depending on the value we choose for $\alpha$, we qualitatively recover a uniform ($\alpha \sim -100$) or a peaked ($\alpha \sim 100$) distribution.

\begin{figure} 
    \centering
    \subfloat[\label{fig:6a}]{
    \includegraphics[width=0.485\textwidth]{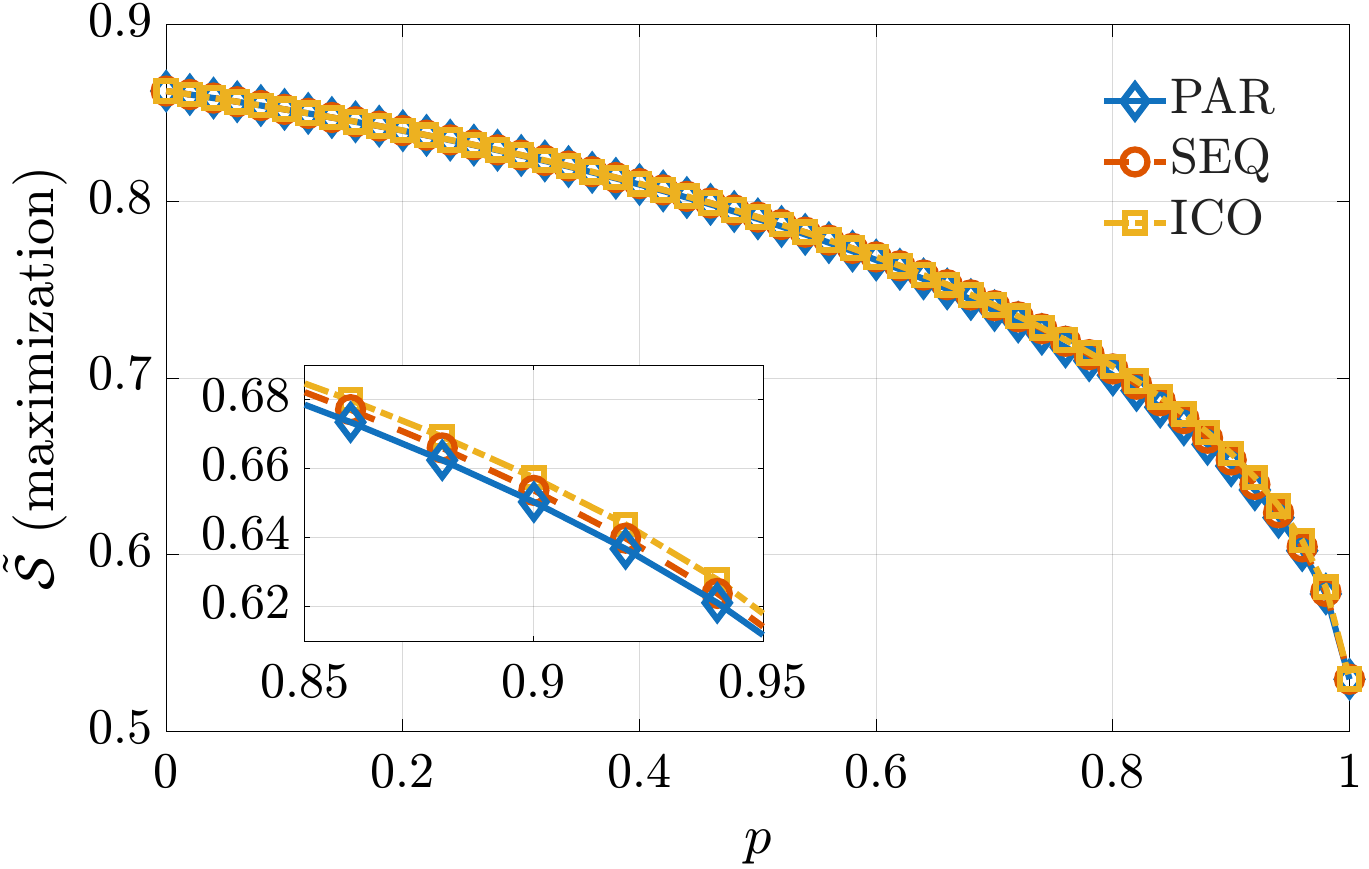}
    }
    \hfill
    \subfloat[\label{fig:6b}]{
    \includegraphics[width=0.49\textwidth]{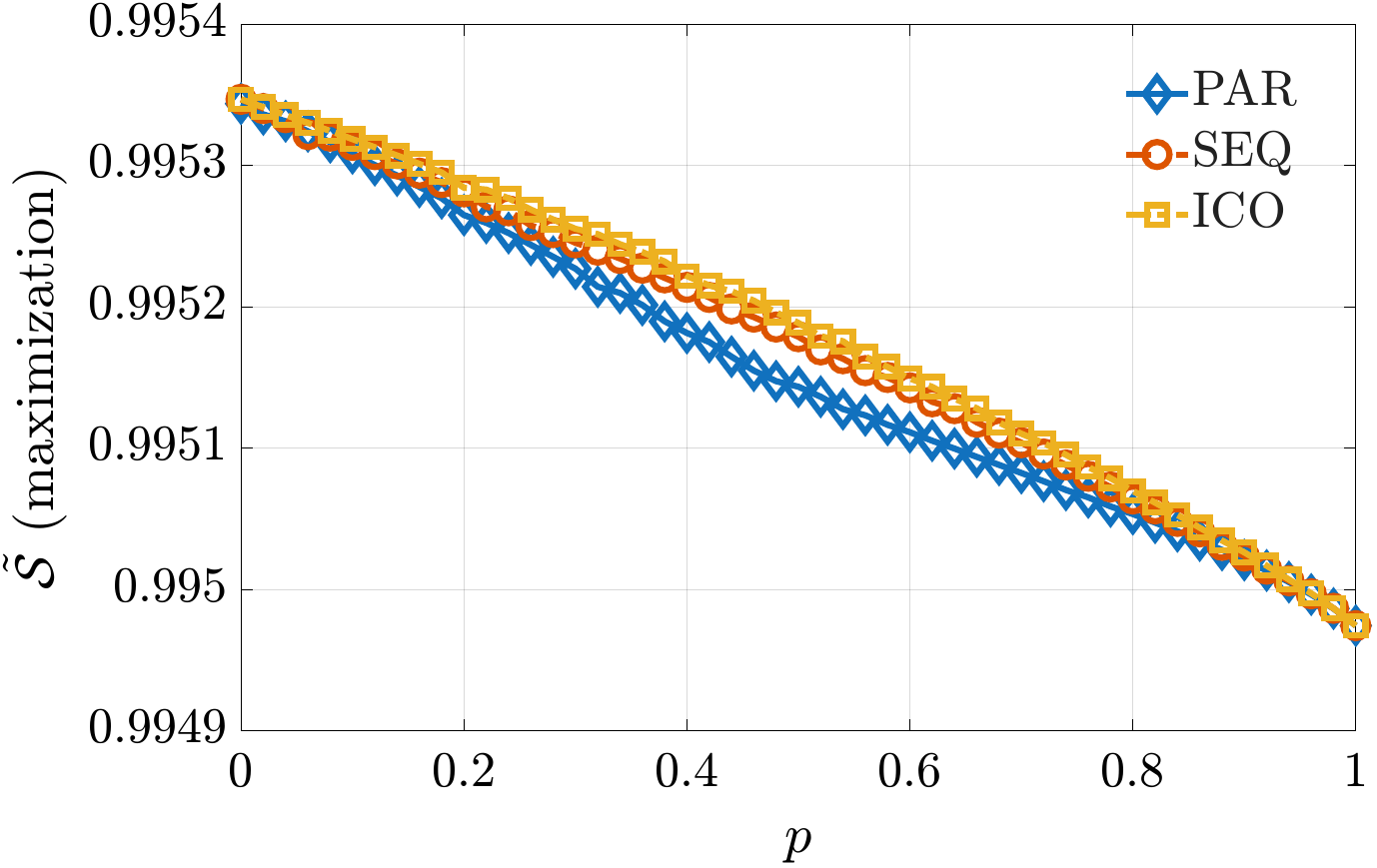}
    }
    \caption{Maximum approximate score $\widetilde{\mathcal{S}}_\mathrm{cos}$ in a phase estimation + AD task for $2$ copies of the channel, both for a {(a)} flat prior ($\alpha = -100$) and {(b)} sharp prior ($\alpha = 100$).}
    \label{fig:phaseestimation+ad}
\end{figure}

In Figure \ref{fig:phaseestimation+ad} we plot the Bayesian score for the different strategies. On the one hand, when considering a flat prior ($\alpha=-100$), we can observe a clear strict hierarchy between strategies whenever $p \in [1/2, 1)$. Note that, again, at the extremal point $p=1$ all strategies are equivalent, for similar reasons that we discussed in the dissipative SU(2) estimation problem. On the other hand, for a peaked prior ($\alpha=100$) we also observe a strict hierarchy between strategies around $p=1/2$, 
but with a lower relative difference than in the flat prior case. 
This reduction is due to the fact that, in the peaked prior regime, the score is heavily dominated by the prior information (see also next section), which is the same for all strategies. Furthermore, note that Figure \ref{fig:6b} presents the same shape as the QFI in Ref.~\cite{PhysRevLett.130.070803}. This is because we are essentially plotting the same figures of merit, as we later discuss. 

\subsection{Connection to the frequentist approach}
To establish such a connection, we have to focus on priors that are sharp, hence pushing the problem into the local estimation regime. In such a scenario, the estimators are expected not to be far from the hypothesis values---at least within the region that the prior is significant---such that $\abs{\hat{\theta}_i - \theta} \ll 1$. Thus, for sharp priors one can write 
\begin{equation}
    c(\theta,\hat{\theta}_i)
    \approx 1 - \dfrac{(\hat{\theta}_i - \theta)^2}{4}.
\end{equation}
Thus, maximizing the average cosine score is, up to an affine rescaling, equivalent to minimizing the mean-squared error (MSE). For a sharp prior, we can write the score as
\begin{align}
    \mathcal{S}_{\rm cos}(T,{\hat \theta}) &= \int \dd{\theta} \sum_{i=1}^N p(\theta) \cos^2 \left( \dfrac{\theta - \hat{\theta}_i}{2} \right) \Tr \left( T_i J_\theta^{\otimes k} \right) \approx \int \dd{\theta} \sum_{i=1}^N p(\theta) \left[ 1 - \dfrac{(\hat{\theta}_i - \theta)^2}{4}\right] \Tr \left( T_i J_\theta^{\otimes k} \right) \\
    & = 1 - \dfrac{1}{4} \mathcal{S}_\mathrm{MSE}(T,{\hat \theta}),
\end{align}
where in the last line we introduced ${\cal S}_{\rm MSE}$, the score based on the MSE:
\begin{align}
    \mathcal{S}_\mathrm{MSE}(T,{\hat \theta}) &\coloneqq  \int \dd{\theta} p (\theta) \sum_{i=1}^{N_\mathrm{O}} p (i | \theta) ( \hat{\theta}_i - \theta )^2 
    =
    \int \dd{\theta} \sum_{i=1}^{N_\mathrm{O}} p(\theta) (\hat{\theta}_i - \theta)^2 \Tr \left( T_i J_\theta^{\otimes k} \right).
\end{align}
For the MSE score, we can use the Van Trees inequality to connect both frequentist and Bayesian approaches ~\cite{VanTrees2004DEMT1, VanTreesBell2007BayesianBounds}:
\begin{equation}\label{VTboundeq}
    \mathcal{S}_\mathrm{MSE}(T,{\hat \theta})  \geq \dfrac{1}{F_0 + \mathbb{E} \left[ \mathcal{F} \left[ p (i | \theta) \right] \right]} \geq \dfrac{1}{F_0 + \mathbb{E} \left[ \mathcal{F}^*(\theta) \right]},
\end{equation}
with $F_0$ being the prior (Fisher) information
\begin{align}
    F_0 &\coloneqq  \int \dd{\theta} p (\theta) \left[ \partial_\theta \log p (\theta) \right]^2, \label{align2}
\end{align}
and $\mathbb{E} \left[ \mathcal{F} \left[ p (i | \theta) \right] \right] = \int \dd{\theta} p (\theta) \mathcal{F} \left[ p (i| \theta) \right]$, with 
\begin{equation}
    \mathcal{F} \left[ p (i| \theta) \right] \coloneqq  \sum_{i=1}^{N_{\rm O}} \dfrac{\left[ \partial_\theta p(i|\theta) \right]^2}{p (i|\theta)},
\end{equation}
being the average classical Fisher information acquired from the protocol leading to the outcome distribution $p (i | \theta)$. Lastly, the quantity ${\cal F}^*(\theta) = \max {\cal F}[p(i|\theta)]$ is the quantum Fisher information, i.e., the maximum of the Fisher information over all allowed states, control operations, and measurements---that is, over allowed testers.
Again, for a sharp prior, we expect that $F_0 \gg \mathbb{E} \left[ \mathcal{F}^*(\theta) \right]$, so we can write
\begin{align}
    \mathcal{S}_\mathrm{MSE}(T,{\hat \theta})  \geq  \dfrac{1}{F_0 + \mathbb{E} \left[ \mathcal{F}^*(\theta) \right]} \approx \dfrac{1}{F_0} - \dfrac{\mathbb{E} \left[ \mathcal{F}^*(\theta) \right]}{F_0^2}.
\end{align}
As $F_0$ depends exclusively on the prior distribution $p(\theta)$, and is specifically independent of the protocol, one would expect that for sharp priors the only protocol-relevant contribution to the score ${\cal S}_{\rm cos}$ comes from the QFI. As such, one would expect any hierarchy that was observed in the frequentist approach to hold also in the Bayesian approach with sharp priors (note that this does not necessarily extend to cost functions that are not reducible to the MSE). We note that, as the prior distribution is very peaked around $\theta_0 = \pi$, we can approximate $\mathbb{E} \left[ \mathcal{F}^*(\theta)\right] \approx \left[ \mathcal{F}^*(\theta_0)\right]$, which gives
\begin{equation}\label{eqd15}
    \mathcal{S}_\mathrm{cos}^*(T,{\hat \theta}) \approx C + \dfrac{\mathcal{F}^*(\theta_0)}{4 F_0^2}, \quad C \coloneqq  1 - \dfrac{1}{4 F_0} \approx 0.995,
\end{equation}
which matches with our numerical simulations in Figure \ref{fig:6b}. 

\end{document}